\documentclass[twocolumn,showpacs,preprintnumbers,amsmath,amssymb,superscriptaddress]{revtex4}

\usepackage{graphicx}
\usepackage{dcolumn}
\usepackage{bm}
\usepackage{longtable}
\newcommand{\eee}{\mbox{\large $e$}}

\newcommand{\Aa}{{\cal A}}

\newcommand{\kk}{{\bf k}}

\newcommand{\rr}{{\bf r}}

\newcommand{\ZZ}{{\cal Z}}

\newcommand{\DD}{{\cal D}}

\newcommand{\Gg}{{\sf G}}

\newcommand{\tr}{\mbox{tr}}
\newcommand{\Tr}{\mbox{Tr}}
\newcommand{\be}{\begin{equation}}
\newcommand{\ee}{\end{equation}}
\newcommand{\ba}{\begin{eqnarray}}
\newcommand{\ea}{\end{eqnarray}}
\newcommand{\bse}{\begin{subequations}}
\newcommand{\ese}{\end{subequations}}
\newcommand{\beq}{\begin{eqnarray}}
\newcommand{\eeq}{\end{eqnarray}}
\newcommand{\ds}{\displaystyle}
\newcommand{\im}{\mbox{Im}}

\newcommand{\ts}{\textstyle}

\newcommand{\Fderiv}[2]{\frac{\delta #1}{\delta #2}}

\newcommand{\w}{\ensuremath{\omega}}

\newcommand{\eq}[2]{\begin{equation}#1\label{#2}\end{equation}}

\begin{document}
\title{Coherent-Potential approximation
for diffusion and wave propagation
in topologically disordered systems}
\author{S. K\"ohler}
 \affiliation{
Institut f\"ur Physik, Universit\"at Mainz, Staudinger Weg 7,
D-55099 Mainz, Germany}
 \affiliation{
Graduate School Materials Science in Mainz, Universit\"at Mainz, Staudinger Weg 9,
D-55099 Mainz, Germany
}
\author{G. Ruocco}
 \affiliation{
Dipartimento di Fisica, Universit\'a di Roma
``La Sapienza'', P'le Aldo Moro 2, I-00185, Roma, Italy}
\author{W. Schirmacher}
 \affiliation{
Institut f\"ur Physik, Universit\"at Mainz, Staudinger Weg 7,
D-55099 Mainz, Germany}
 \affiliation{
Dipartimento di Fisica, Universit\'a di Roma
``La Sapienza'', P'le Aldo Moro 2, I-00185, Roma, Italy}

\begin{abstract}
Using Gaussian integral transform techniques borrowed
from functional-integral field theory and the replica trick
we derive a version of the coherent-potential
approximation (CPA) suited for describing
($i$) the diffusive (hopping)
motion of classical particles in a random
environment and ($ii$) the vibrational properties of
materials with spatially fluctuating elastic coefficients
in topologically disordered materials.
The effective medium in the present version of the CPA
is not a lattice but a homogeneous and isotropic medium,
representing an amorphous material on a mesoscopic scale.
The transition from a frequency-independent to a frequency-dependent
diffusivity (conductivity) is shown to
correspond to the boson peak in the vibrational
model. The anomalous regimes 
above the crossover are governed by a complex, frequency-dependent
self energy. The boson peak is shown to be stronger for non-Gaussian
disorder than for Gaussian disorder. 
We demonstrate that the low-frequency non-analyticity of the
off-lattice version of the CPA leads to the correct
long-time tails of the velocity autocorrelation function in the hopping
problem and to low-frequency Rayleigh scattering in the wave problem.
Furthermore we show
that the present version of the CPA
is capable to treat the percolative aspects of hopping
transport adequately.
\pacs{05.40.-a 63.50.-x}
\end{abstract}
\maketitle

\section{Introduction}
The coherent-potential approximation (CPA) is a very successful
mean-field theory for treating quenched disorder 
\cite{velicky68,Yonezawa73,Eliot74,Janis89,Vollhardt10}
and strong correlations \cite{dmt1,dmt2,dmt3,kakehashi02} in quantum
systems. It also proved useful for describing classical
diffusion \cite{summerfield81,webman81,odagakilax81} and vibrational
excitations \cite{schirm98,taraskin01} in disordered systems.
However, until now, The CPA has never been applied to the study
of topologically disordered systems, as all its versions involve
a regular crystalline lattice with additional quenched disorder.
As it is, indeed, desirable
to describe excitations of quenched-disordered systems with
no crystalline disorder (amorphous metals, glasses, amorphous semiconductors)
we present in this paper a field-theoretic derivation of the CPA
for the two formally equivalent problems of diffusion and
wave propagation in a topologically disordered environment.
Our effective medium
is not a lattice, but a homogenous
and isotropic
continuum. Section 2 is devoted to the mathematically equivalent
problems of diffusion and scalar waves with spatially fluctuating
diffusivity/elastic modulus. For this model our version of the
CPA is derived and solved. From the numerical solutions it is demonstrated
that the existence of a
$dc$ - $ac$ crossover in the diffusion problem corresponds to
the presence of an excess of vibrational states (boson peak)
\cite{schirm06,schirm07,marruzzo_forthcoming1,marruzzo_forthcoming2,schirm11,schirm13}
in the wave propagation problem. It is shown that the height
of the boson peak is unrestricted for certain non-Gaussian
distributions (inverse-power law distribution, log-normal distribution)
but restricted for a box-shaped and a Gaussian distribution.
For the non-Gaussian distributions
a scaling relation
between the height of the boson peak and its frequency position is shown
to hold within the CPA.
For weak
disorder the CPA is shown to reduce to the self-consistent Born approximation
(SCBA), which proved very successful to describe the anomalous
vibrational properties of disordered solids 
\cite{schirm06,schirm07,marruzzo_forthcoming1,marruzzo_forthcoming2,schirm11,schirm13}.
In section 3 the CPA for vector displacements
in an elastic continuum with fluctuating shear modulus is presented
(heterogeneous-elasticity theory
\cite{schirm06,schirm07,marruzzo_forthcoming1,marruzzo_forthcoming2,schirm11,schirm13}). The vector theory is shown to
possess the same boson-peak features as the scalar theory.

\section{Diffusion and scalar vibrations in a disordered environment}
\subsection{Models and mathematic correspondence}
Let us start our discussion with considering
hopping transport of electrons or ions in a disordered
semiconductor, or -- equivalently -- among the impurities
of a doped crystalline semiconductor
\cite{bottger81,efros84}. Such a motion of
particles between specific sites $i,j$ in a disordered
material can be described by a master equation for the
probability $n_i(t)$ for being at site $i$ at time $t$
\be\label{discrete}
\frac{d}{dt}n_i(t)=-\sum_{j\neq i}W_{ij}\big(n_i(t)-n_j(t)\big)
\ee
where the transition (hopping) probabilities per unit
time $W_{ij}$ can depend on the distance $r_{ij}=|\rr_i-\rr_j|$
and/or on an energy barrier $E_{ij}$ between
the sites $i$ and $j$ (representing a symmetrized version
of diffusion in a landscape of states with disordered
local energies with phonon-assisted transitions between
them \cite{bottger81,efros84}). The distance dependence is supposed
to fall off exponentially, so that we do not have a long-range model.

It has been shown recently \cite{ganter10} that
by a coarse-graining procedure
such a transport equation can be transformed to
a diffusion equation on a {\it mesoscopic} scale
\be\label{eqmo1}
\frac{\partial}{\partial t}n(\rr,t)
=\nabla D(\rr)\nabla n(\rr,t).
\ee
with a spatially fluctuating diffusion coefficient
$D(\rr)$, which is supposed to be a random variable in the three-dimensional
space with a suitable distribution density $P[D(\rr)]$.
Performing the
Laplace transform
\mbox{$n(\rr,s)=\int_0^\infty dt e^{-st}n(\rr,t)$}
with $s=i\omega+\epsilon$ we obtain from (\ref{eqmo1}) the diffusion
equation in frequency space (disregarding the $t\!=\!0$ term)
\be\label{eqmo1a}
s n(\rr,s)
=\nabla D(\rr)\nabla n(\rr,s).
\ee

On a {\it macroscopic} scale the (quenched) disorder is known to lead
to a diffusion equation with a space-independent but frequency-dependent
complex diffusivity $D(s)$
\be\label{eqmo1b}
s n(\rr,s)
=D(s)\nabla^2 n(\rr,s).
\ee
$D(s)$ is the Laplace-Transform of the velocity autocorrolation
function $Z(t)$ of the moving particle. If the particle carries a charge $q$,
this quantity is related to the complex, frequency-dependent
conductivity $\sigma(s)$ by the Nernst-Einstein relation
\be
\sigma(s)=n_\mu q^2 D(s)\, ,
\ee
where $n_\mu\equiv\partial n/\partial \mu$ is the derivative of the
number of carriers with respect to the chemical potential. In degenerate
quantum systems this quantity is equal to the density of electronic
states at the Fermi level, in classical systems
$n_\mu=n/k_BT$, where $T$ is the temperature.

Let us now consider a topologically disordered mass-spring system
in which the masses (which we suppose to be equal to unity) are connected
by distance-dependent force constants, which we call
$K_{ij}$. The corresponding
equation of motion for the scalar displacements of the
masses at point $i$ is
\be\label{discrete2}
\frac{d^2}{dt^2}u_i(t)=-\sum_{j\neq i}K_{ij}\big(u_i(t)-u_j(t)\big)
\ee
The same coarse-graining procedure, which leads from (\ref{discrete}) to
(\ref{eqmo1}) produces the following stochastic wave equation
\be\label{eqwa1}
\frac{\partial^2}{\partial t^2}u(\rr,t)
=\nabla K(\rr)\nabla u(\rr,t).
\ee
or in frequency space
\be\label{eqwa1a}
s^2u(\rr,s)\equiv\tilde s u(\rr,\tilde s)=\nabla
K(\rr)\nabla u(\rr,\tilde s).
\ee
where now $K(\rr)\equiv v(\rr)^2$ has the meaning of a space dependent
modulus, which is equal to the square of the
wave velocity $v$.
$K(\rr)$  is again a random variable in the three-dimensional
space and can be identified with $D(\rr)$ in the diffusion problem.

On a macroscopic scale one deals with an equation of motion in frequency
space for scalar wave amplitudes $u(\rr,z)$
with a frequency-dependent, complex sound velocity $v(\tilde s)$
\be\label{eqwa1b}
\tilde s u(\rr,\tilde s)=
K(\tilde s)\nabla^2 u(\rr,\tilde s).
\ee
Here $\tilde s=s^2=-\omega^2+i\tilde \epsilon$ and
$K(\tilde s)=v^2(\tilde s)$ is a complex, frequency-dependent elastic modulus, which
is equal to the square of a complex wave velocity. As in optics
the imaginary part of $v(\tilde s)$, $v''(\omega)$ is related to the 
disorder-induced mean-free
path of the waves, $\ell(\omega)$ and to the sound-attenuation
coefficient $\Gamma(\omega)$ by 
\cite{ganter10}
\be
\frac{1}{\ell(\omega)}
=\frac{2\omega v''(\omega)}{|v(\tilde s)|^2}
=\frac{1}{2|v(\tilde s)|}\Gamma(\omega)
\ee
In a quenched-disordered system, i.e. a medium with either spatially
fluctuating density or elastic modulus the sound attenuation
exhibits Rayleigh scattering 
\cite{RayleighBlueSky,ganter10}
\be
\Gamma(\omega)\propto \omega^4\qquad \mbox{for} \qquad \omega\rightarrow 0
\ee
This means that both $v(\tilde s)$ and $K(\tilde s)$ have a contribution, which
varies as $\tilde s^{3/2}$ for small frequencies. 
If we now mathematically identify the 
diffusion
coefficient $D(s)$ with $K(\tilde s)$ we conclude that
$D(s)$ has a low-frequency non-analytic $s^{3/2}$ contribution
\be\label{nonana}
\Delta D(s)=D(s)-D(0)\propto s^{3/2}
\ee
which
becomes $\Delta D(s)\propto s^{d/2}$ in $d$ dimensions.
This non-analytic asymptotics has been proven to hold for
any quenched-disordered system governed by equations of
motion of the form 
(\ref{discrete}),
(\ref{eqmo1}),
(\ref{discrete2}), and
(\ref{eqwa1}).
As $D(s)$ is the
Laplace transform of the velocity autocorrelation function $Z(t)$ one
obtains the non-analytic long-time asymptotics (``long-time tail'')
$Z(t)\propto -t^{(d+2)/2}$. This long-time-tail property, which has been
known already for some time
\cite{ernst71,%
machta84,ernst84},
is obviously equivalent to Rayleigh scattering via the
correspondence $D(s)\leftrightarrow K(\tilde s)$ \cite{ganter10,ganter11}.
The analogy, in fact, goes further: In a quenched-disordered system (which
is the subject-matter of the present work) it is known that a
cross-over happens between a frequency-independent diffusivity
and a strong frequency dependence, which can be parametrized
as $D(s)\propto s^x$ with $x\approx 0.8$ \cite{long82,jonscher97,dyre00}.
This disorder-induced diffusivity transforms under the
corresondence $D(s)\leftrightarrow K(\tilde s)$ to
disorder-induced anomalous frequency dependence of the elastic
modulus, the onset of which corresponds to the boson
peak \cite{schiwagener92,schiwagener93}. This correspondence
will be discussed in more detail below.

\subsection{Derivation of the CPA}
We now 
consider the mathematically equivalent problems
Eqs. (\ref{eqmo1a}) and (\ref{eqwa1a}),
identifying 
the quantities $D(s)$, $K(\tilde s)$ and $s$, $\tilde s$.
We
define \footnote{We use the conventional
bra-ket formalism of quantum mechanics, i.e.
$<\rr|u^\alpha>=u^\alpha(\rr)$, etc. .}
\be
(s-\nabla D(\rr)\nabla )\delta(\rr-\rr')\equiv
<\rr'|\Aa[D(\rr)]|\rr>
\ee

The Green's function corresponding to Eq. (\ref{eqmo1}) is given by the inverse
matrix element of $\Aa$:
\be
\Gg(\rr,\rr')=<\rr'|\Aa^{-1}|\rr>\, .
\ee

Applying standard methods in replica field theory \cite{john83}
we represent the Greens function as a functional integral
over mutually complex conjugate fields $u(\rr)^\alpha$ and $\bar u(\rr)^\alpha$
present in $\alpha=1, \dots, n$ replicas
of the system as follows: 
\ba
\Gg(\rr,\rr')
&=&\prod\limits_{\alpha=1}^n\int
\DD[\bar u^\alpha(\rr),
u^\alpha(\rr)]
\bar u^1(\rr)
u^1(\rr')\nonumber\\
&&\times\,\eee^{\ts -\sum\limits_\alpha<u^\alpha|\Aa|u^\alpha>}\\
&=&\frac{\delta}{
\delta J^{(1)}(\rr,\rr')
}\ZZ[J(\rr,\rr')]
\ea
with the generating functional
\ba\label{genfunc}
\ZZ[J(\rr,\rr')]&=&
\prod\limits_{\alpha=1}^n\int
\DD[\bar u^\alpha(\rr)]
\DD[u^\alpha(\rr)]\,
\eee^{\ts -\sum\limits_\alpha<u^\alpha|\Aa|u^\alpha>}\nonumber\\
&&\eee^{\ts -\sum\limits_\alpha<u^{(\alpha)}|J^{(\alpha)}|u^{(\alpha)}>}
\ea
and the source-field $J^{\alpha}$.
By an integration by part the 
matrix elements of the
inverse Green operator $\Aa$ can be written as
\be
<u^{\alpha}|\Aa[D]|u^{\alpha}>=
\int d^3\rr \bigg(
s|u^\alpha(\rr)|^2+
D(\rr)|\nabla u^\alpha(\rr)|^2\bigg)
\ee
We now apply the Fadeev-Popov procedure \cite{belitz97}, which consists of the replacement of the fluctuating
diffusivity $D(\rr)$ by a complex auxiliary field $Q^{(\alpha)}(\rr,s)$ with the help of
a delta functional, which,
in turn, is represented by another auxiliary
field $\Lambda^{(\alpha)}(\rr,s)$ 

\ba
\ZZ[J]&=&
\int\mathcal{D}[u,\bar{u}]\,\int\mathcal{D}[Q]\,\eee^{-<u|\Aa[Q]-J|u>}\delta[D-Q]\nonumber\\
&=&\int\mathcal{D}[u,\bar u]\,\mathcal{D}[Q,\Lambda]\,\eee^{-<\phi|\Aa[Q]-J|\phi>}\eee^{<\Lambda|D-Q>}\nonumber\\
&=&\int\mathcal{D}[Q,\Lambda]\,\eee^{-\Tr\{\,\ln[\,\Aa[Q]-J\,]\,\}}\eee^{<\Lambda|D-Q>}\label{AuxFields}
\ea

where we have suppressed the replica indices for
brevity. The third equality in Eq. (\ref{AuxFields}) follows
from integrating out the displacement fields
$\bar u^\alpha$ and $u^\alpha$.
In order to proceed further we devise
another coarse-graining procedure.

We tile the total space into $N_c$ cells of (approximate) volume
$V_c=V/N_c$, where $V=L^3$ \footnote{We work in $d=3$ dimensions
throughout this paper, although the analysis
is not limited to this dimension.} is
the total volume. This could just be done by means of a cubic grid.
However, in order to avoid any relation to a crystalline
lattice we think, instead, of a
a Voronoi tessellation
around midpoints of a closed-packed hard-sphere structure.
This gives $V_c=L_c^3=(\pi/6)\eta_c d_c^3$, where $\eta_c\approx 0.56$
is the close-packed packing fraction, resulting in
$L_c\approx 0.66 d_c$. Within a cell 
with label $i$ we
replace the diffusivity by their average in each cell and assume
that a diffusion equation
\be\label{eqmo2}
\frac{\partial}{\partial t}n(\rr,t)
=\nabla D_i\nabla n(\rr,t).
\ee
holds within a cell with label $i$. We now assume that
the random numbers $D_i$ are independent of each other,
i.e. the joint distribution density is assumed to factorize as
$P(D_1 \dots D_{N_c})=\prod\limits_i \,p(D_i)$.

Our assumption of independent fluctuations of the quantities
$D_i$ implies that the size of the cells $L_c$ must be larger
or at least equal to the correlation length $\xi$ of the
diffusivity fluctuations, which is defined by
\be
\xi^3=\frac{1}{\langle D^2\rangle}
\int d^3\rr \langle D(\rr+\rr_0)D(\rr_0)\rangle\, .
\ee
Correspondingly we confine the $\kk$ summations in the subsequent
analysis to remain below a cutoff \mbox{$|\kk|<k_\xi=\nu/\xi$}, 
where $\nu$ is an adjustable number of the order of 
1.

Within our model $D(\rr)$ is now a piecewise constant function in real space and the same should hold for 
the auxiliary fields $Q$ and $\Lambda$,
which are now labeled as
$Q^{(\alpha)}_i$,
$\Lambda^{(\alpha)}_i$.
Using this the scalar product, which appears in the
exponential in Eq.  
(\ref{AuxFields}) can be written as:

\be
<\Lambda|D-Q>
=\frac{V_c}{V}\sum_{\alpha}\sum_i\Lambda_i^{(\alpha)}(\bm{r})\left(D_i^{(\alpha)}-Q_i^{(\alpha)}\right)
\ee
We now start to evaluate the configurational average.
Due to the Fadeev-Popov transformation the only
term to be averaged over is the term 
$e^{<\Lambda|D-Q>}$.

Assuming that all the $N_c$ coarse-graining cubes behave the 
same on average and using that the 
individual cubes are not correlated, we can write

\begin{eqnarray}
&\left\langle e^{<\Lambda|D-Q>}\right\rangle
=\prod_\alpha \prod_i
\left\langle \eee^{\ts\frac{V_c}{V}\Lambda_i^{(\alpha)}(D_i^{(\alpha)}-Q_i^{(\alpha)})}\right\rangle_i \nonumber&\\
&=\eee^{\sum_{\alpha}\frac{V}{V_c}\ln\left(\left\langle \exp\left[\,-\frac{V_c}{V}\Lambda_i^{(\alpha)}(D_i^{(\alpha)}-Q_i^{(\alpha)})\,\right] \,\right\rangle_i\,\right)} \label{avCGVolume}&
\end{eqnarray}

Note that the two occurring volume ratios do not cancel each other due to the average inside the logarithm.
Using (\ref{avCGVolume}) the generating
functional (\ref{AuxFields}) can be written as

\be\label{final}
\ZZ[\tilde J]=\int\mathcal{D}[Q,\Lambda]\, \eee^{\ts-S_{\text{eff}}[Q,\Lambda,\tilde J]}
\ee
where we have now replaced the source field
$J^\alpha(\rr,\rr')$ by translational-invariant source
field $\tilde J(\rr-\rr')$ which is  not supposed
to depend on the replica index $\alpha$.
The effective action takes the form
\ba\label{skalphonseff}
S_{\text{eff}}[Q,\Lambda,\tilde J]&=&\Tr\{\,\ln\big(\Aa[Q]-\tilde J\big)\}\\
&&-\displaystyle\sum_{\alpha=1}^n\frac{V}{V_c}\ln\left(\left\langle \eee^{-\frac{V_c}{V}\Lambda_i^{(\alpha)}(D_i^{(\alpha)}-Q_i^{(\alpha)})} \,\right\rangle_i\,\right)\nonumber
\ea
Since the factor $\frac{V}{V_c}$ in the effective action (\ref{skalphonseff}) is much larger than unity a saddle point approximation can be employed to evaluate the integral in (\ref{final}). In general this factor will scale as 
\be
\frac{V}{V_C}=\left(\frac{L}{\xi}\right)^d\stackrel{d\to\infty}{\longrightarrow}\infty
\ee
Accordingly the CPA becomes exact for $d\rightarrow \infty$
\cite{Vollhardt10}. 

We now assume that the fields $Q$ and $\Lambda$ are
replica independent.

\begin{subequations}
\be
 S_{\text{eff}}[Q,\Lambda,0]=n\,S_{\text{eff}}^\prime(\{Q_i\},\{\Lambda_i\})\\
\ee
\ba\label{scalPhon_EffAction}
S_{\text{eff}}^\prime(\{Q_i\},\{\Lambda_i\})&=&\tr\{\,\ln[\,\widetilde{A}(Q)\,]\,\}\\
&&-\sum_{i}\ln\left(\left\langle \eee^{-\frac{V_c}{V}\Lambda_i(D_i-Q_i)} \,\right\rangle_i\,\right)\, ,\nonumber
\ea
\end{subequations}
where ``tr'' now means a trace whithout the replica indices.
 
The saddle point is determined by the equations

\begin{subequations}
\begin{eqnarray}
 \left.\frac{\partial S^\prime_{\text{eff}}}{\partial Q_i}\right|_{Q_i=Q_{i,S}}&=&0,\,\forall i\\\label{calPhon_QEq}
 \left.\frac{\partial S^\prime_{\text{eff}}}{\partial\Lambda_i}\right|_{\Lambda_i=\Lambda_{i,S}}&=&0,\,\forall i\label{calPhon_LEq}
\end{eqnarray}
\end{subequations}

The derivative with respect to $\Lambda$ is easily performed and yields:
\begin{eqnarray}
 0&=&\frac{\left\langle-\frac{V_c}{V}\Lambda_{i,s}(D_i-Q_{i,s})\eee^{-\frac{V_c}{V}\Lambda_i(D_i-Q_{i,s})}\right\rangle_i}{\left\langle \eee^{-\frac{V_c}{V}\Lambda_{i,s}(D_i-Q_{i,s})}\right\rangle_i} \nonumber\\
\Rightarrow\qquad0&=&\left\langle\frac{D_i-Q_{i,s}}{\exp[\frac{V_c}{V}\Lambda_{i,s}(D_i-Q_{i,s})]}\right\rangle_i\label{saddleb}
\end{eqnarray}

Since $\frac{V_c}{V}\ll1$ the exponential in the denominator can be expanded to first order \footnote{If the exponential in the denominator of
Eq. (\ref{saddleb}) would remain in the numerator and then expanded,
we would obtain the self-consistent Born approximation (see below)}:

\be\label{scalPhon_QSaddle}
0=\left\langle\frac{D_i-Q_{i,s}}{1+\frac{V_c}{V}(D_i-Q_{i,s})\Lambda_{i,s}}\right\rangle_i
\ee

The second saddle point equation gives

\ba\label{ScalPhon_LambdaTraceLog}
\left.\frac{\partial\,\tr\{\,\ln[\,\widetilde{A}(Q)\,]\,\}}{\partial Q_i}\right|_{Q_i=Q_{i,s}}&=&\nonumber
\frac{\frac{V_c}{V}\Lambda_{i,s}\left\langle\eee^{-\frac{V_c}{V}\Lambda_i(D_i-Q_{i,s}} \right\rangle_i}{\left\langle\eee^{-\frac{V_c}{V}\Lambda_{i,s}(D_i-Q_{i,s}} \right\rangle_i}\\&=&\frac{V_c}{V}\Lambda_i
\ea

The left-hand side can be evaluated under the assumption that the saddle point field $Q_S$ is constant in space
\[Q_{i,s}\equiv Q,\,\forall i\]
This corresponds to the introduction of an effective homogeneous medium. In this medium (\ref{ScalPhon_LambdaTraceLog}) becomes:

\be\label{ScalPhonLambdaTr}
\frac{V_c}{V}\Lambda =\frac{V_c}{V}\frac{\partial}{\partial Q}\tr\ln[A_{\text{eff}}]=\frac{V_c}{V}\sum_{\bm{k}}\frac{k^2}{s+Qk^2}
\ee

In the second step the effective-medium operator 

\be\label{ScalPhon_effMediumOp}
A_{\text{eff}}(\bm{k},\widetilde{\bm{k}})=(s+Q k^2)\delta_{\bm{k}\widetilde{\bm{k}}}
\ee

was defined. From
this representation
one can see
that in the CPA the following holds:

\be\label{CPA_AvGreen}
\langle G\rangle(\bm{k},\widetilde{\bm{k}},s) =\frac{1}{s+Q k^2}\delta_{\bm{k}\widetilde{\bm{k}}}=\left\langle\frac{1}{s+Dk^2}\right\rangle\delta_{\bm{k}\widetilde{\bm{k}}}
\ee

under the assumption that the averaged system exhibits translational invariance.  This equation expresses the averaged Green's function in terms of the Green's function of a homogeneous medium, where the spatially fluctuating diffusivity is replaced by the self energy $Q$, which, however, is now frequency dependent:
The space dependence due to the disorder has been transformed
to a disorder-induced frequency dependence.

From (\ref{ScalPhonLambdaTr}) it follows that if $Q$ is homogeneous
in space,
the same holds for $\Lambda$.
Defining a new field 
\mbox{$\widetilde\Lambda=3V_c/\widetilde{\nu} V\Lambda$} 
with $\widetilde{\nu}=\nu^3/2\pi^2$ 
and performing the summation in (\ref{ScalPhonLambdaTr}) with a cutoff $|\bm{k}|<k_\xi$ the CPA equations become:

\begin{subequations}
 \begin{eqnarray}
\label{cpa1a}
0&=&\left\langle\frac{D_i-Q(s)}{1+\frac{\ts \widetilde \nu}{\ts 3}\big[D_i-Q(s)\big]\widetilde\Lambda(s)}\right\rangle_i\\
\label{cpa1b}
\widetilde\Lambda(s)&=&
\frac{3}{k_\xi^3}\int_0^{k_\xi}dk k^2\frac{k^2}{s+k^2Q(s)}\\
&=&\frac{1}{Q(s)}[1-sG(s)]\nonumber
\end{eqnarray}
with the local Green's function
\be\label{green}
G(s)=\frac{3}{k_\xi^3}\int_0^{k_\xi}dk k^2 \frac{1}{s+k^2Q(s)}
\ee
We call $\widetilde\Lambda(s)$ the susceptibility function, because
it is proportional to the local dynamic susceptibility
of the diffusing particle.

The CPA equation (\ref{cpa1a}) can be cast into the following
equivalent forms
 \begin{eqnarray}
\label{cpa1c}
1&=&\left\langle\frac{1}{1+\frac{\ts \widetilde \nu}{\ts 3}\big[D_i-Q(s)\big]\widetilde\Lambda(s)}\right\rangle_i\\
\label{cpa1d}
Q(s)&=&\left\langle\frac{D_i}{1+\frac{\ts \widetilde \nu}{\ts 3}\big[D_i-Q(s)\big]\widetilde\Lambda(s)}\right\rangle_i
\end{eqnarray}
\end{subequations}
It is worthwhile to note that the $k$ integral in
Eq. (\ref{cpa1b}) for the susceptibility function can be carried
out analytically. The diffusion pole in the denominator of the
integrand produces a $\widetilde\Lambda(s)=
\widetilde\Lambda(0)\,+\,\mbox{const}\times s^{2/3}$
low-frequency asymptotics, which is inherited by the
function $Q(s)$. So, independent of the type of disorder, we
obtain a correct long-time behaviour for $Z(t)$ and Rayleigh
scattering for the wave problem. Because 
for the vibrational problem in the $s\rightarrow 0$
limit the disorder scattering is suppressed by the Rayleigh
frequency dependence, the CPA expression 
for the imaginary part of $Q(z)$ can be shown to
reduce to the Born approximation in agreement with the previous
derivations \cite{RayleighBlueSky,ganter10}.

In the $dc$ limit $s=0$ we have
$\widetilde\Lambda(0)=1/Q(s)$, so we
obtain from Eq. (\ref{cpa1d})
\begin{subequations}
\be\label{cpa1f}
Q(0)=\left\langle
\frac{D_i}{1-\frac{\widetilde\nu}{3}+\frac{\widetilde\nu}{3}\frac{D_i}{3Q(0)}}
\right\rangle
\ee
(From now on we suppress the index $i$, which indicates the
average over $p(D_i)$.)
In the case $Q(0)\neq 0$ (which is not trivial, see the paragraph
on percolation) one can divide by $3Q(0)/\widetilde\nu$ to obtain
\be\label{cpa1g}
\frac{\widetilde\nu}{3}=\left\langle
\frac{1}{1+\big(\frac{3}{\widetilde\nu}-1\big)\frac{Q(0)}{D_i}}
\right\rangle
\ee
\end{subequations}

\subsection{Relation with previous effective-medium theories}
\subsubsection{Lattice CPA}
The standard 2-site coherent-potential approximation
for the hopping problem of Eq. (\ref{discrete}) on a lattice
is \cite{odagakilax81,summerfield81,webman81}
\be\label{cpa1}
\left\langle\frac{W_{ij}-\Gamma(s)}{1+(W_{ij}-\Gamma(s)
\frac{2}{Z\Gamma(s)}
(1-sG_{ii}(s)}
\right\rangle=0
\ee
Here $\Gamma(s)$ is the effective frequency-dependent hopping rate
and $G_{ii}(s)$ is the local Green's function of the effective
medium, which whithin this theory is a simple-cubic lattice with
coordination number $Z=2d=6$. The lattice Green's function has the form
\be
G_{ii}(s)
=\sum_{\kk\in BZ}\frac{1}{s + \Gamma(s)f(\kk)}
\ee
where the sum goes over the first Brillouin zone (BZ), and
\be
f(\kk)=
6-2[
\cos(k_xa)+
\cos(k_ya)+
\cos(k_za)]
\ee
with $a$ being the lattice constant.
Defining the local susceptibility function
\be
\Lambda_{ii}(s)=\frac{1}{\Gamma(s)}\bigg(1-sG_{ii}(s)\bigg)
=\sum_{\kk\in BZ}\frac{f(\kk)}{s + \Gamma(s)f(\kk)}
\ee
the CPA equation (\ref{cpa1}) takes the form
\be\label{cpa2}
\left\langle\frac{W_{ij}-\Gamma(s)}{1+(W_{ij}-\Gamma(s)
\frac{1}{3}\Lambda_{ii}(s)}
\right\rangle=0
\ee
Now we take the continuum limit by replacing the BZ $\kk$ summation
by $\sum_{\kk}\rightarrow \frac{3}{k_\xi^3}\int_0^{k_\xi}dk$ and
$f(\kk)$ by its low-wavenumber limit $k^2a^2$. If we now
define
the local diffusivities by $D_i=W_{ij}a^2$, the effective-medium
diffusivity by $Q(s)=\Gamma(s)a^2$ and the continuum susceptibility
function by $\widetilde\Lambda(s)=\Lambda_{ii}(s)/a^2$ we arrive
at the continuum-off-lattice CPA result (\ref{cpa1a}), provided
we take $\widetilde\nu=1$. In order to be consistent with the
continuum limit of the
lattice CPA one may take always this value.
On the other hand, one can also use this value such
that the CPA percolation threshold
$p_c^{\rm CPA}=\widetilde\nu/3$
(see below) agrees to the continuum percolation threshold $p_c$
of a certain topology
\cite{butcher,Mov2}.
\subsubsection{Self-consistent Born approximation, SCBA}
If one takes Gaussian disorder for the local diffusivity one
can perform the disorder average over the generating functional
(\ref{genfunc}) exactly, which leads to an interacting effective
field theory with the variance of $D(\rr)$ as coupling constant.
Taking apart this interaction by a Hubbard-Stratonovich approximation
and then performing a saddle-point approximation (assuming a
small relative variance of $D(\rr)$)
one arrives
at the {\it self-consistent Born approximation} for the
scalar problem \cite{maurer02,maurer04,MaurerFELO}. We can, however
recover the SCBA from the CPA in the following way.
Defining $D_0$ to be the average of the fluctuating diffusivities
and defining the quantities
$Q(s)=D_0-\Sigma(s)$,
$D_i=D_0-\Delta_i$, we obtain from (\ref{cpa1a}) the two
(equivalent)
CPA equations
\begin{subequations}
\ba\label{cpa3a}
0&=&\left\langle\frac{\Delta_i-\Sigma(s)}{1-\frac{\widetilde \nu}{3}(\Delta_i-\Sigma(s))\widetilde\Lambda(s)}\right\rangle_i\\
\label{cpa3b}
\Sigma(s)&=&\left\langle\frac{\Delta_i}{1-\frac{\widetilde \nu}{3}(\Delta_i-\Sigma(s))\widetilde\Lambda(s)}\right\rangle_i\label{ScalPhon_finalQ}
\ea
\end{subequations}
We now expand the interior of the average in (\ref{cpa3b}) with
respect to $\Delta_i-\Sigma(s)$ to lowest nonvanishing
order (respecting $\langle\Delta_i\rangle=0$) we obtain
\be\label{scba}
\Sigma(s)=\langle D_i^2\rangle\frac{\widetilde\nu}{3}\widetilde\Lambda(s)
\ee
which is the SCBA for the scalar problem. As indicated already above, the SCBA
can also be obtained from the saddle-point equation
(\ref{saddleb}) by putting the exponential not into the denominator
but into the numerator and then expand with respect to the
small number $V_c/V$ to first order. Because then only the first two cumulants
of the distribution of the $D_i$ enter, this corresponds to 
assuming Gaussian disorder.

So we recover the SCBA from the CPA in the
Gaussian and weak-disorder limit.
\subsubsection{Network effective-medium approximation, EMA}
The CPA-like effective-medium treatment of the impedances of a
heterogeneous medium or network date back to
Bruggeman  \cite{bruggemann35} and Landauer \cite{landauer52}.
For a disordered $Z$-fold coordinated network of
fluctuating
conductances $g_i$ the expression for the
effective-medium conductance $g_m$
is \cite{kirkpatrick73}
\be
0=\left\langle\frac{g_m-g_i}{g_i+\left(\frac{Z}{2}-1\right)g_m}
\right\rangle
\ee
It has been generalized for the $ac$ problem \cite{dyre94,dyre00},
setting $Z/2=d$
\be
0=\left\langle\frac{g_m(s)-g_i}{g_i+(d-1)g_m(s)+ds}
\right\rangle
\ee
which can be re-arranged as
\be
0=\left\langle
\frac{g-g_m(s)}{1+\big[g-g_m(s)\big]\frac{1}{d}\frac{1}{g_m(s)+s}}
\right\rangle
\ee
which has - for $d=3$ - the same form as the CPA equation (\ref{cpa1a})
with 
\be\label{lambdaema}
\widetilde \Lambda(s)_{\rm EMA}=\frac{1}{s+Q(s)}
\ee
This EMA
has the same analytical structure as
EMA versions derived earlier
in the literature
\cite{Mov2,Andersen,butcher}. While these $ac$  effective-medium
theories describe rather nicely measured hopping conductivity
data they violate the non-analyticity requirement (\ref{nonana}).
As stated above,
in CPA the function $\widetilde\Lambda(s)$ has a contribution,
which
varies as $s^{3/2}$, whereas $\widetilde \Lambda(s)_{\rm EMA}$
does not, as can be clearly seen from
Eq. (\ref{lambdaema}).
\subsection{CPA results for the $dc$ diffusivity}
\subsubsection{Percolation}
In order to treat the percolation problem, which can be
considered as the continuum version of the Lorentz problem
\cite{machta84,ernst84,hofling07,spanner11}, we assume a distribution of local diffusivities
of the form
\be
P(D_i)=p\delta(D_i-D_0)+(1-p)\delta(D_i)
\ee
where $p$ is the volume fraction in which the diffusivity
is non-zero.
Inserting this into Eq. (\ref{cpa1f}) one obtains
\be\label{perc1}
Q(0)
\frac{\widetilde\nu}{3}=Q(0)
\frac{1}{1+\big(\frac{3}{\widetilde\nu}-1\big)\frac{Q(0)}{D_0}}
\ee
which has, like Eq. (\ref{cpa1f}) always the trivial solution
$Q(0)=0$.
For the case $Q(0) \neq 0$ we obtain from Eq. (\ref{perc1})
\be\label{perc2}
Q(0)=\frac{2}{3}D_0(p-p_c)/(1-p_c)
\ee
with $p_c=\frac{\widetilde\nu}{3}$. 
For $p<p_c$ the trivial solution of (\ref{perc1}), 
$Q(0)=0$ takes over.

\subsubsection{Activated diffusion}
In this class of models the diffusion of a particle
is considered to take place by jumps over barriers
of height $E_i$ with a certain distribution $P(\epsilon)$
(``random barrier model'' \cite{dyre00}).
So we write 
\be\label{activated}
D_i=D_0e^{E_i/k_BT}
\ee
 and parametrize
the $dc$ diffusivity as $(\frac{3}{\widetilde\nu}-1)Q(0)=D_0e^{E_a/k_BT}$. Then
Eq. (\ref{cpa1g}) takes the form
\be\label{barr}
\frac{\widetilde\nu}{3}=\int dE P(E)\frac{1}{e^{(E-E_a)/k_BT}+1}
\ee
In the low-temperature limit the Fermi function becomes a step
function and we obtain
\be\label{barr1}
\frac{\widetilde\nu}{3}=p_c=\int_0^{E_a}dEP(E)
\ee
which
means that the parameter $E_a$ becomes temperature independent.
From this follows that in the random-barrier problem the
$dc$ diffusivity is {\it always} of Arrhenius form, as
observed frequently in fast-ion conducting glasses
\cite{SolidStateIonics}.

It is worthwhile to point out that (\ref{barr1}) corresponds to the
so-called percolation construction for obtaining the $dc$ conductivity
of a disordered hopping-conduction network \cite{efros84,dyre00}. 
It has been nicely demonstrated recently that the low-temperature
physics of the random-barrier model is essentially
percolation physics \cite{dyre08}.
\subsubsection{Variable-range hopping}
A rather widely investigated type of carrier diffusion in disordered
materials is that of electrons performing phonon-assisted tunneling
transitions between localized states (``hopping transport'' \cite{efros84}).
Understanding the mechanims of
electronic hopping transport has been shown recently to be of extreme
importance for devising organic light-emitting diodes (OLEDs) 
\cite{bobbert}. Here we show that by the CPA one recovers
the classical results of Mott \cite{mott} 
and Efros, Shklovski{\u i} \cite{efros84}
for variable-range hopping.

\begin{figure}
\vspace{1ex}\includegraphics[width=.45\textwidth]{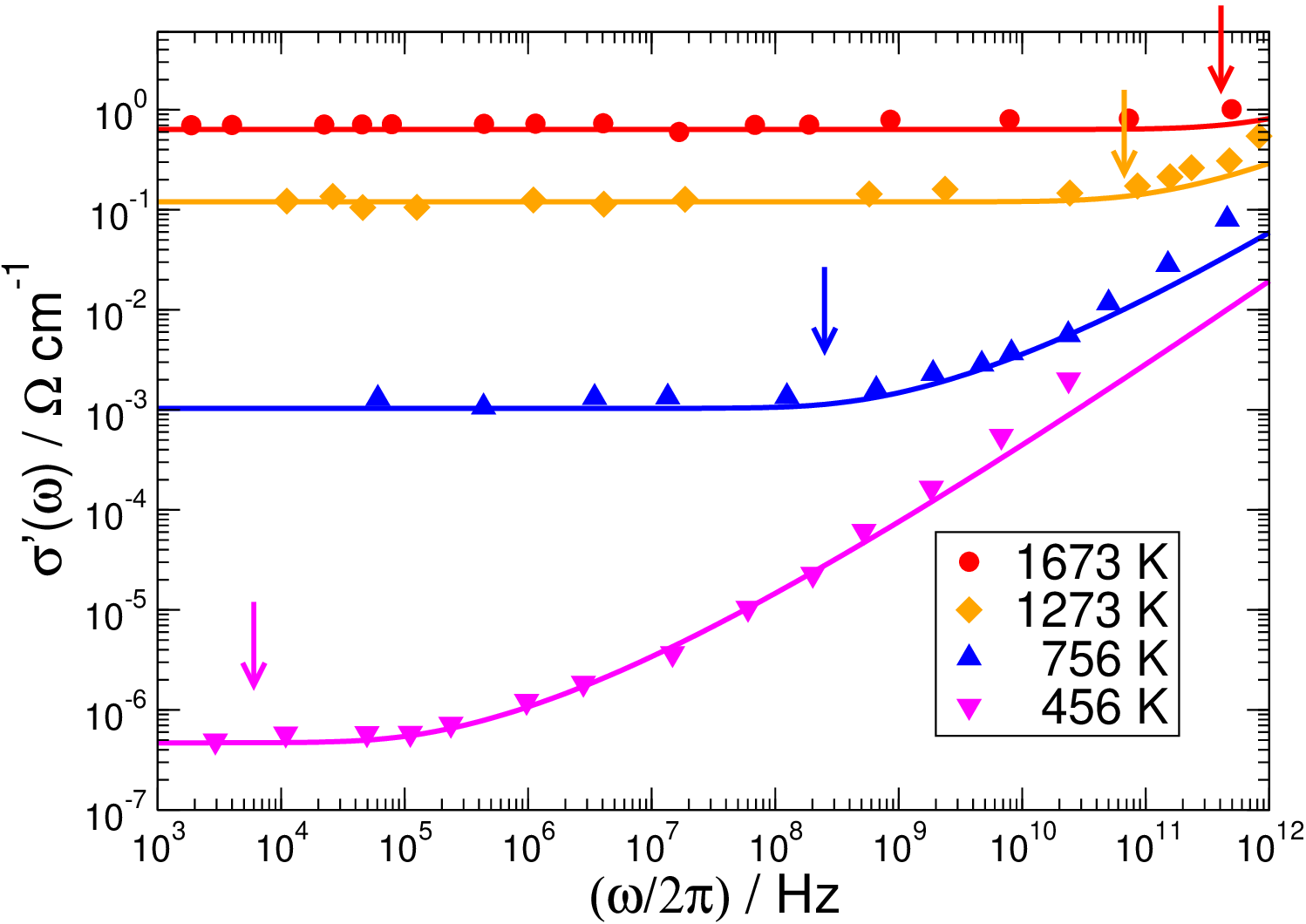}
\caption{
Comparison of ionic ac conductivity data of sodium trisilicate  glass \cite{WongAngel}
with the CPA prediction for activated hopping with a constant-barrier distribution. We use units in which $D_0=k_\xi=1$. The arrows indicate the boson-peak
positions $\omega_{\rm BP}\equiv\tilde\omega_{\rm BP}^2$ of Fig. \ref{skalboson}.
}
\label{angel}
\end{figure}

The local diffusivity depends on an activation barrier $E$
and a characteristic hopping distance $r$ with distributions
$P(E)$ and $P(r)$
\be
D_i=D_0e^{-\alpha r-\beta E}
\ee 
where $e^{-\alpha r}$ is the tunneling factor and
$\beta=1/k_BT$.
Depending on the density of localized electronic states
near the Fermi energy
$P(E)$ is either considered to be
constant (Mott hopping) or proportional to $E^2$
(Coulomb-gap, Efros-Shklovski hopping). The distribution
of sites is
\be
P(r)=\frac{1}{Z}4\pi\rho r^2\theta(R-r)=\frac{3}{R^3}r^2\theta(R-r)
\ee
where $Z=\frac{4}{3}4\pi\rho R^3$ is the number of adjacent sites within
a given radius $R$.
If we parametrize 
the $dc$ diffusivity as $(\frac{1}{p_c}-1)Q(0)=D_0e^{-\xi}$,
we obtain from Eq. (\ref{cpa1g})
\be
p_c=\int dE P(E)\int dr P(r)
\frac{1}{e^{\alpha r+\beta E-\xi}+1}
\ee
In the low-temperature limit the Fermi function becomes again a
step function, and, by means of integrations by part
one obtains the famous results
$\ln Q(0)\propto -(T_0/T)^{1/4}$ (Mott hopping)
$\ln Q(0)\propto -(T_0/T)^{1/2}$ (Efros-Sklovskii hopping).
Again, these results are equivalent to the percolation construction
\cite{efros84}.
\subsection{CPA results for the ac diffusivity}
We consider activated transport of the form (\ref{activated})
with a constant barrier distribution
\be\label{constantbarrier}
P(E)=\frac{1}{E^*}\qquad 0\leq E\leq E^*
\ee
which is equivalent to an inverse-Power distribution \mbox{for $D$}
\be
P(D)=\frac{1}{\ln{\mu/\sigma}}\frac{1}{D} \qquad \mu\leq D\leq\sigma
\ee
with $\sigma=D_0$ and $\mu=D_0e^{-\beta E^*}$.

\begin{figure}
\vspace{1ex}\includegraphics[width=.45\textwidth]{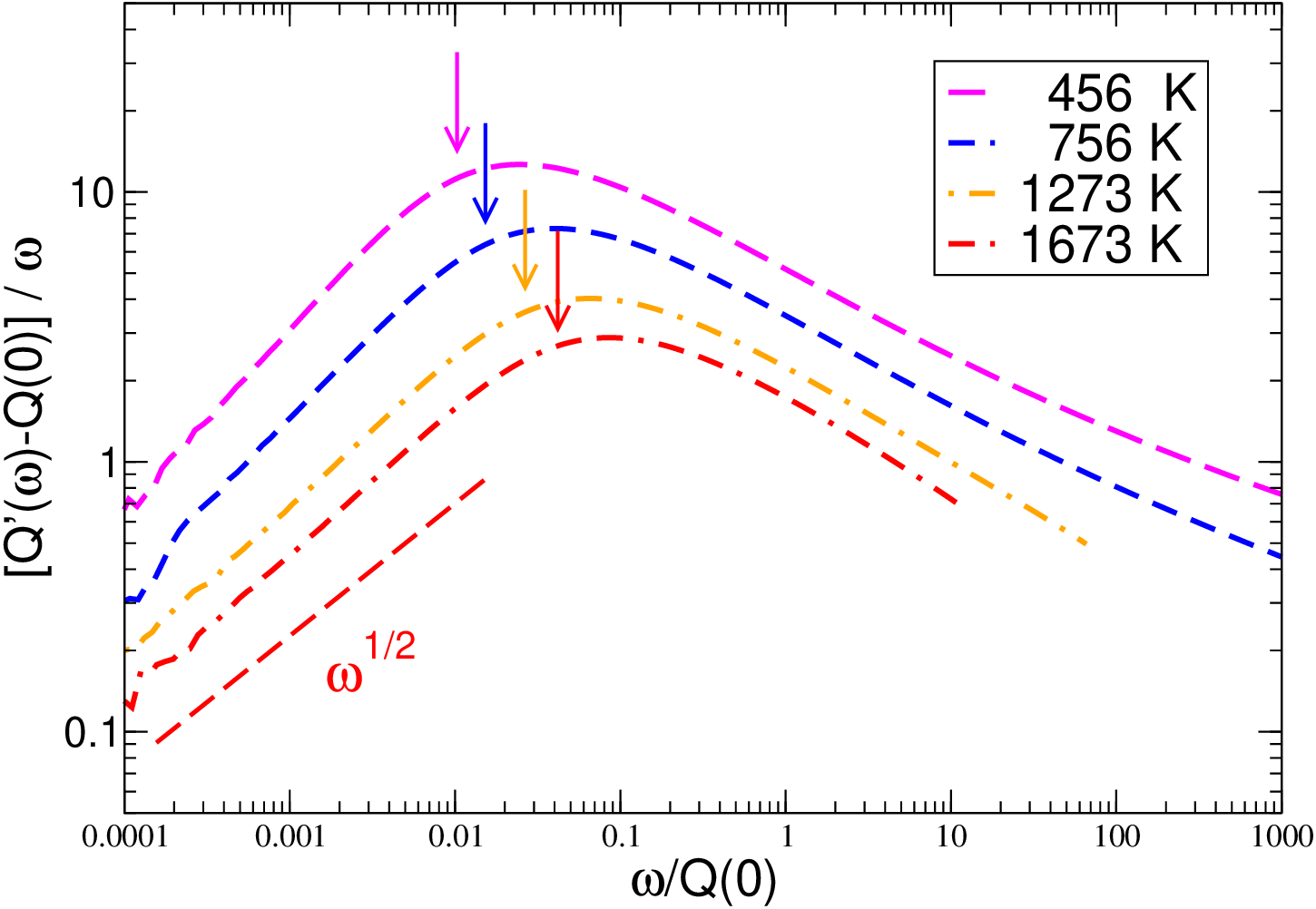}
\caption{\label{losstheo} 
Loss function
$\big[Q'(\omega)-Q(0)\big]/\omega\propto
\big[\sigma(\omega)-\sigma(0)\big]/\omega$
calculated from the CPA curves in Fig. \ref{angel} corresponding to
the temperatures \mbox{456 K,} 756 K and 1273 K. Below the $dc$ -$ac$ crossover
the Rayleigh-type non-analyticity is visible.}
\end{figure}

In Fig. \ref{angel} we show ac conductivity data collected from the literature
over a very wide range of frequencies by Wong and Angell \cite{WongAngel}
together with the CPA prediction for the constant-barrier model (\ref{constantbarrier}).
The only input is the measured $dc$ activation energy of 
$E_a$ = 75 KJ/mole
\footnote{Note that $E_a$ is {\it not} equal to the
maximum barrier height $E^*$ but much smaller and
is determined by the self consistent Eq. (\ref{barr}).}
However, it should be
made clear that the old EMA theories 
\cite{Mov2,Andersen,butcher,dyre85,dyre00a} are also able to produce such a fit. The 
difference to the CPA can be seen from the 
``loss function'' \cite{long82,Lossfunction}
$[\sigma(\omega)-\sigma(0)]/\omega\propto 
[Q'(\omega)-1]/\omega$,
which is shown in Fig. \ref{losstheo}. At low frequencies
this function behaves as $\omega^{1/2}$ due to the Rayleigh-type
non-analyticity. The old EMA theories do not exhibit
this non-analyticity.
In the figure this behaviour
is demonstrated. There is experimental \cite{Lossfunction} and
simulational \cite{pasveer06} evidence for this low-frequency non-analyticity
of hopping transport.

\subsection{Model calculations for the scalar phonon problem}
We shall now exploit the
mathematical correspondence
$i\omega\leftrightarrow -\omega^2$, $Q(s)\equiv D(s)\leftrightarrow 
Q(\tilde s)\equiv K(\tilde s)$
and, correspondingly, discuss the vibrational anomalies
induced by the quenched disorder, as given in CPA.
Similar discussions have already published in the literature
\cite{schiwagener92}, where it was pointed out that
the boson peak (BP) in the vibrational problem corresponds to
the $dc$ - $ac$ crossover in the diffusion problem.

As shown in the last subsection, the activated-diffusion
model with a constant barrier distribution
corresponds to an inverse-power law distribution for
the local diffusivities. For the corresponding local
moduli $K$ we re-write this distribution
\be\label{inverse}
P(K)=\frac{1}{\ln{\mu/\sigma}}\frac{1}{K} \qquad \mu\leq K\leq\sigma
\ee
Other possible distributions, namely a uniform distribution, a Gaussian
distribution, truncated at $K=0$ and a log-normal distribution
are detailed in table \ref{DistTable}.
The density of states (DOS) of the scalar phonons can be calculated as
\be
\label{ScalPhon_DOS}g(\omega)=\frac{2\omega}{\pi} \im\big\{G(\tilde s)\big\}
\ee 
where $G(\tilde s)$ is given by Eq. (\ref{green}), and we identify
the correlation cutoff $k_\xi$ with the Debye cutoff $k_D$.

\begin{table}[h]
\begin{center}
 \begin{tabular}{lc}\hline\hline
  Name & P(K) \\\hline
&\\
  Uniform & ${\ds(\sigma-\mu)^{-1},\,K\in[\mu,\sigma]}$\\
&\\
  Truncated Gaussian& ${\ds\sqrt{\frac{2}{\pi \sigma}}\frac{\exp[-(K - \mu)^2/(2 \sigma)]}{1 + \text{Erf}(\mu/\sqrt{2 \sigma})} }$\\
&\\
  Power Law&${\ds(K\ln[\frac{\sigma}{\mu}])^{-1},\,K\in[\mu,\sigma]}$\\
&\\
  Log-Normal&${\ds\frac{1}{\sqrt{2 \pi \sigma}} \frac{1}{K} e^{\ts-\ln[\frac{K}{\mu}]^2/(2 \sigma)}}$\\
&\\
\hline\hline
 \end{tabular}
\end{center}
 \caption[Distribution functions and their cumulants]{\label{DistTable}Probability distributions used to model the disorder.}
\end{table}

As mentioned above, the CPA makes it possible to describe
highly disordered systems, i.e. systems in which the variance
of the spatial fluctuations of the quantity of interest
exceeds the square of its average.
In order to quantify the strength of the disorder, 
we define a disorder parameter as the ratio between the variance and the squared mean of the disorder distribution
$\gamma=\langle K^2\rangle/\langle K\rangle^2$
For the truncated Gaussian and the uniform distribution $\gamma$ has an upper bound. For the uniform distribution a maximum disorder strength of $\gamma=\frac{1}{3}$ can be reached, for the truncated Gaussian this limit is $\gamma=\frac{\pi}{2}-1$. Thus two of the four distributions can only model medium to weak disorder. On the other hand, the inverse-power and log-normal distributions
have no upper bound of the disorder parameter. In particular, for the
inverse-power distribution with $\mu=\sigma e^{-\beta E^*}$ the
relation $\gamma=\beta E^*/2$ holds.

\begin{figure}
	\begin{center}
\vspace{1ex}\includegraphics[width=0.45\textwidth]{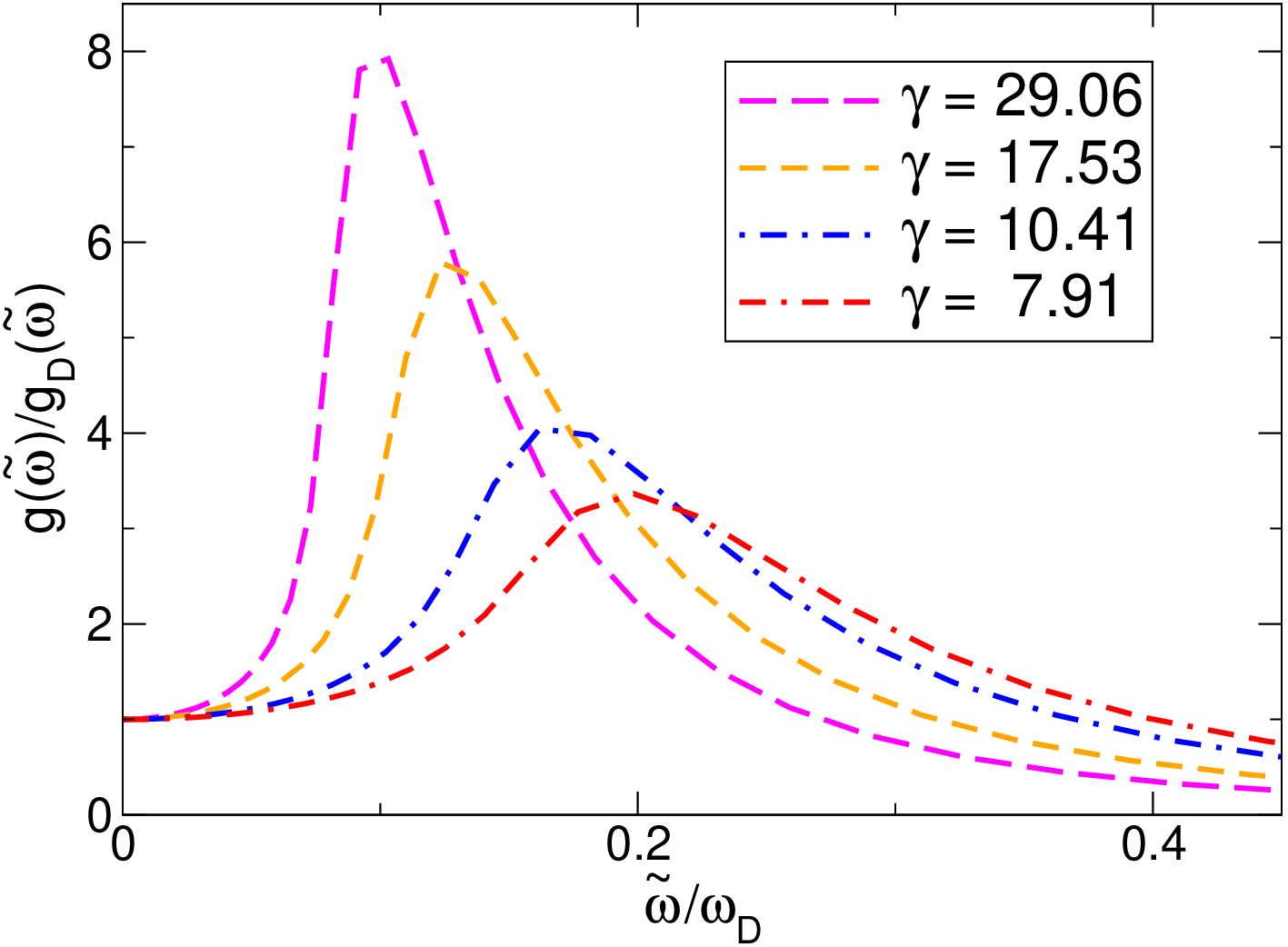}
	\caption{\label{skalboson} Reduced density of states
$g(\tilde\omega)/g_D(\tilde\omega)$
for inverse-power law disorder with the same disorder parameters
as in Figs. \ref{angel} and \ref{losstheo}. As the disorder increases so does the height of the boson peak, it is also shifted to lower frequencies.}
	\end{center}
\end{figure}

In Fig. \ref{skalboson} we show the so-called reduced DOS
$g(\tilde\omega)/%
g_D(\tilde\omega)$, calculated for the inverse-power-law distribution. 
We use units, in which
$k_\xi=k_D=1$ and $K_0=1$.
$g_D(\omega)=3\omega^2/\omega_D^3$ is the Debye DOS,
and $\omega_D=\sqrt{Q(0)}$ is the Debye frequency.

The disorder parameters $\gamma=\beta E^*/2$ have been chosen to agree to those in
the conductivity calculations of Fig. 
\ref{angel}. The boson peaks 
shown in Fig. \ref{skalboson} increase with increasing disorder, while its position
decreases. The positions $\tilde\omega_{\rm BP}$
of the boson peaks are indicated in Figs.
\ref{angel} and \ref{losstheo} 
via the correspondence $\omega_{\rm BP}\leftrightarrow\tilde\omega_{\rm BP}^2$.
It is clearly seen that the boson peak marks the onset of the disorder-induced
frequency dependence of the diffusivity ($\equiv$ conductivity). Via the correspondence
$D(\omega)\leftrightarrow K(\tilde \omega)=v(\tilde\omega)^2$ this means that the
boson peak 
marks the frequency dependence of the sound velocity in the disordered vibration model.
This is in agreement with earlier conclusions from effective-medium
calculations using the EMA \cite{schiwagener92,schiwagener93} 
and the SCBA \cite{schirm06,schirm07,marruzzo_forthcoming1,marruzzo_forthcoming2,schirm11,schirm13}. 

\begin{figure}
\begin{center}
\vspace{1ex}\includegraphics[width=.4\textwidth]{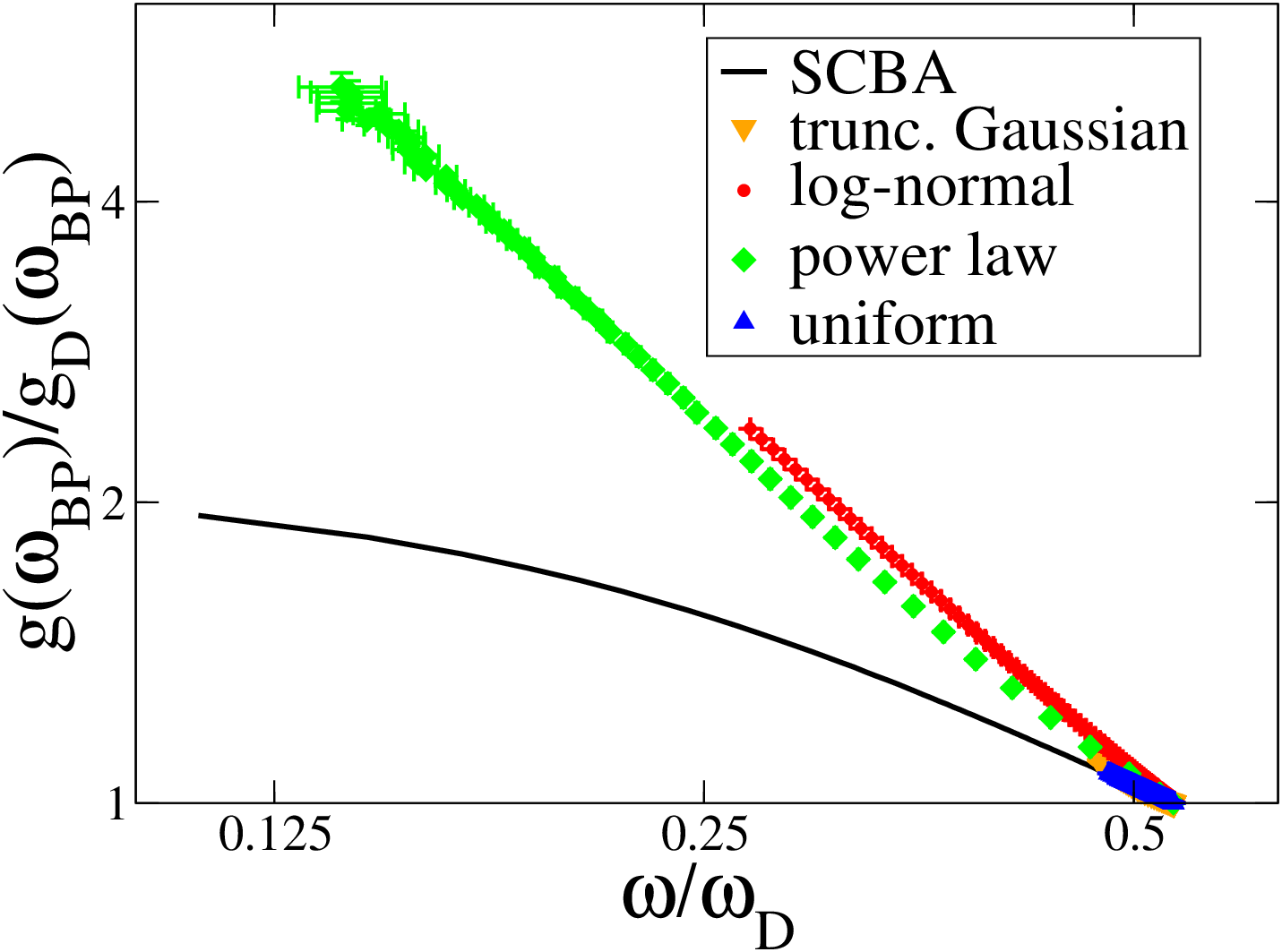}
\caption{\label{scalingplot} Relation between boson peak height and position for the CPA solutions of different distributions and disorder strength for $k_\xi=k_D$.}
\end{center}
\end{figure}

\begin{figure}
\begin{center}
\vspace{1ex}\includegraphics[width=.45\textwidth]{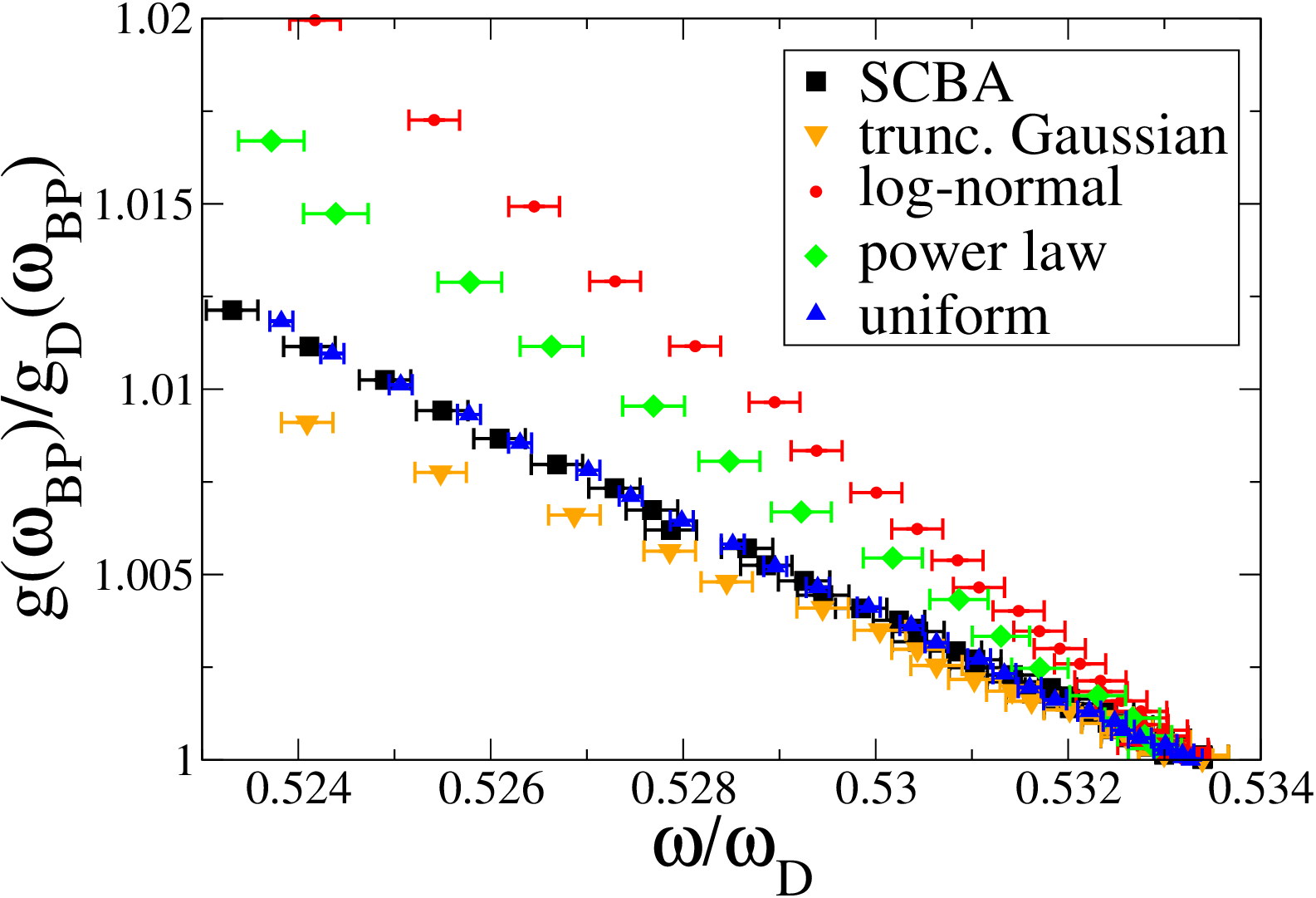}
\caption{\label{scalingplot_closeup} Close up of the low disorder region of Fig. \ref{scalingplot}. Note that in the low-disorder limit 
$\omega_{\rm BP}\rightarrow $ 0.5333, independently of 
of the chosen disorder distribution.}
\end{center}
\end{figure}

In Fig. \ref{scalingplot} the boson-peak height is plotted
against its frequency position for all four distributions 
considered. We include also the prediction of the
self-consistent Born approximation (SCBA). Fig. 
\ref{scalingplot_closeup} shows a close up of the
small-disorder region. 

The down-shift and reinforcement of the boson peak with increasing disorder
has been shown in the SCBA-based earlier treatments of vibrational
anomalies \cite{schirm13} to result from the disorder-induced level repulsion, which
is present in the anomalous frequency regime above the boson peak. This
level repulsion, which is typical for random-matrix spectra, results from the
absence of symmetries on the microscopic scale.

From our CPA calculations, displayed in Fig. \ref{scalingplot} it follows that
the height of the BP scales with its frequency position $\omega_{\rm BP}$ as
$\frac{g(\omega_{BP})}{g_D(\w_{\rm BP})}\propto(\omega_{\rm BP})^{c}.$
with $c=1.25$. It is suggestive that this scaling should be related
to the power-law frequency dependence of the diffusivity in
the mathematically equivalent diffusion problem, but we did not
find a way to prove this.


\section{CPA for heterogeneous-elasticity theory}
\subsection{Model}
We start with the equations of motion of elasticity theory
in frequency space, formulated in terms of the stress tensor
$\sigma_{ij}=\rho\tilde\sigma_{ij}$ ($\rho$ is the mass density)
\cite{LandauLifshitz}:

\be\label{model1}
-\omega^2u_i(\rr,\omega)
-\sum_j\partial_j\tilde\sigma_{ij}(\rr,\omega)
\equiv
\sum_j A_{ij}u_j(\rr,\omega)\, ,
\ee
where $u_i(\rr,\omega)$ are the Cartesian components of the
displacement field.
We consider an elastic medium in which the elastic shear modulus
may fluctuate in space. If the system is assumed to
be still isotropic, the stress tensor can be represented
in two ways
\begin{subequations}
\ba
\label{model2a}
\tilde\sigma_{ij}&=&\frac{1}{\rho}\sigma_{ij}=\tilde \lambda\delta_{ij}\tr\{\epsilon\}
+2\tilde G(\rr)\epsilon_{ij}\\
\label{model2b}
&=&\frac{1}{\rho}\sigma_{ij}=\tilde K\delta_{ij}\tr\{\epsilon\}
+2\tilde G(\rr)\widehat\epsilon_{ij}
\ea
\end{subequations}
Here $\lambda=\rho\tilde\lambda=K+\frac{2}{3}G$ is
the longitudinal Lam\'e modulus,
$K=\rho\tilde K$ is the bulk modulus and 
$G=\rho\tilde G$ the shear modulus.
$\epsilon_{ij}$ is the strain tensor
($\partial_j\equiv\partial/\partial x_j$)
\be\label{model3}
\epsilon_{ij}=1/2(\partial_iu_j+\partial_ju_i)
\ee
and $\widehat\epsilon_{ij}$ the traceless strain tensor
\be\label{model4}
\widehat\epsilon_{ij}=\epsilon_{ij}-\frac{1}{3}\delta_{ij}\tr\{\epsilon\}
\ee
($\partial_j\equiv\partial/\partial x_j$).
The spatial fluctuations of the shear modulus
can be modeled in two ways: In what we call {\it Model I.} \cite{schirm06,schirm07} the longitudinal Lam\'e 
modulus $\lambda$
is assumed to be constant
(referring to the representation (\ref{model2a}) ),
in {\it Model II.} \cite{marruzzo_forthcoming1,marruzzo_forthcoming2} the $K$ modulus
(referring to the representation (\ref{model2b}) )
is assumed to be constant. As can be shown easily, in model I. the macroscopic
longitudinal Lam\'e modulus is frequency independent, in model II. the
macroscopic bulk modulus is frequency independent \cite{schirm06,marruzzo_forthcoming1,marruzzo_forthcoming2}.

\subsection{Derivation of the CPA for heterogeneous elasticity}
For model I. the matrix $A_{ij}$ takes the explicit form
\ba\label{matrix1}
A_{ij}&=&-\omega^2\delta_{ij}-\tilde \lambda\partial_i\partial_j\\
&-&\bigg(\partial_j\tilde G(\rr)\partial_i+\delta_{ij}
\sum_\ell\partial_\ell\tilde G(\rr)\partial_\ell\bigg)\, ,\nonumber
\ea
for model II. we have
\ba\label{matrix2}
A_{ij}&=&-\omega^2\delta_{ij}-\tilde K\partial_i\partial_j\\
+\frac{2}{3}\partial_i G(\rr)\partial_j
&-&\bigg(\partial_j\tilde G(\rr)\partial_i+\delta_{ij}
\sum_\ell\partial_\ell\tilde G(\rr)\partial_\ell\bigg)\nonumber
\ea
The Green matrix $G=A^{-1}$
is represented as
\ba
\Gg(\rr,\rr')_{ij}
&=&\prod\limits_{\alpha=1}^n\int
\DD[\bar u^\alpha_\ell(\rr),
u^\alpha_m(\rr)]
\bar u^1_i(\rr)
u^1_j(\rr')\nonumber\\
&&\times\,\eee^{\ts -\sum\limits_\alpha<u^\alpha_\ell|\Aa|u^\alpha_m>}\\
&=&\frac{\delta}{
\delta J^{(1)}_{ij}(\rr,\rr')
}\ZZ[J(\rr,\rr')]
\ea
with the generating functional
\ba\label{genfuncv}
\ZZ[J(\rr,\rr')]&=&
\prod\limits_{\alpha=1}^n\int
\DD[\bar u^\alpha_\ell(\rr)
u^\alpha_m(\rr)]\,
\eee^{\ts -\sum\limits_\alpha<u^\alpha_\ell|\Aa|u^\alpha_m>}\nonumber\\
&&\eee^{\ts -\sum\limits_\alpha<u^\alpha_\ell|J^\alpha_{\ell m}|u^\alpha_m>}
\ea
and the source-field matrix $J^{\alpha}$.
By an integration by part one arrives
at the following representation of the action
\ba
\sum_{ij}
<u^{\alpha}_i|\Aa[G]|u^{\alpha}_j>&=&
\int d^3\rr \bigg(
\tilde s\sum_i
|u_i^\alpha(\rr)|^2
+\frac{1}{2}\tilde K\tr\{\epsilon^\alpha(\rr)\}^2\nonumber\\
&+&
G(\rr)\sum_{ij}|\widehat\epsilon^\alpha_{ij}(\rr)|^2\bigg)
\ea
Applying again
the Fadeev-Popov procedure and performing all the steps we have
done before, we arrive at an effective
action, which looks similar to that of the
diffusion problem (\ref{skalphonseff})
\ba\label{scalPhon_Seffv}
S_{\text{eff}}[Q,\Lambda,\tilde J]&=&\Tr\{\,\ln\big(\Aa[Q]-\tilde J\big)\}\\
&&-\displaystyle\sum_{\alpha=1}^n\frac{V}{V_c}\ln\left(\left\langle \eee^{-\frac{V_c}{V}\Lambda_i^{(\alpha)}(D_i^{(\alpha)}-Q_i^{(\alpha)})} \,\right\rangle_i\,\right)\nonumber
\ea
But now the trace operation goes also over the Cartesian indices.

The homogeneus Matrix $A_{\rm eff}[Q]$ is both diagonal in the Cartesian 
indices and with respect to the $\kk$ vectors. The diagonal elements
are $1/G_L(\kk,\tilde s)$ and twice $1/G_T(\kk,\tilde s)$, which are given by
\begin{subequations}
\be
1/G_L(\kk,\tilde s)=\tilde s+k^2[\tilde \lambda+2Q(\tilde s)]
\quad\mbox{Model I}
\ee
\be
1/G_L(\kk,\tilde s)=\tilde s+k^2[\tilde K+\frac{4}{3}Q(\tilde s)]
\quad\mbox{Model II}
\ee
\be
1/G_T(\kk,\tilde s)=\tilde s+k^2Q(\tilde s)
\ee
\end{subequations}
The saddle-point equations, followed by the expansion of
the exponential in the denominator leads to the CPA equations

\begin{subequations}
\be\label{cpa1av}
0=\left\langle\frac{G_i-Q(\tilde s)}{1+\frac{\widetilde \nu}{3}(G_i-Q(\tilde s))\Lambda(\tilde s)}\right\rangle_i
\ee
The susceptibility functions for the models $\alpha=I,II$ take the form
\be
\Lambda_\alpha(\tilde s)=
\frac{3}{k_\xi^3}\int_0^{k_\xi}dk k^4\bigg(q_\alpha G_L(k,\tilde s)+2G_T(k,\tilde s)\bigg)
\ee
\end{subequations}
with $q_I=2$ and $q_{II}=4/3$.
The density of states is calculated from the well-known formula
\be
g(\omega)=\im\bigg\{
\frac{2\omega}{3\pi}\bigg(G_L(\tilde s)+2G_T(\tilde s)\bigg)
\bigg\}
\ee
where we have introduced the local longitudinal and transverse
functions (identifying again $\xi$ with $k_D$)
\be
G_{L,T}(\tilde s)=\frac{3}{k_D^3}\int_0^{k_D}dkG_{L,T}(k,\tilde s)
\ee
We can write the susceptibility function $\Lambda_\alpha(\tilde s)$ as follows
\ba
\Lambda_\alpha(\tilde s)&=&\frac{1-\tilde sG_T(\tilde s)}{Q(\tilde s)}
+q_\alpha\frac{1-\tilde sG_L(\tilde s)}{p_\alpha+q_\alpha Q(\tilde s)}\\
&=&\widetilde\Lambda(\tilde s)
\bigg(
1+\frac{q_\alpha Q(\tilde s)}{p_\alpha +q_\alpha Q(\tilde s)}
\frac{1-\tilde sG_L(\tilde s)}{1-sG_T(\tilde s}
\bigg)\nonumber
\ea
with $p_I=\tilde\lambda$ and $p_{II}=\tilde K$.
$\widetilde\Lambda(\tilde s)$, which is the transverse
local susceptibility, is the same mathematical function of
$Q$ as
the susceptibility
function of Eq. (\ref{cpa1b}) for the scalar phonon problem.
Because the function inside
the big brackets is only weakly frequency dependent
and the density of states is dominated by the transverse Green's function
all the results derived and presented for the scalar phonon problem
hold also for the vector phonon problem. 

\begin{figure}
\begin{center}
\vspace{1ex}\includegraphics[width=.45\textwidth]{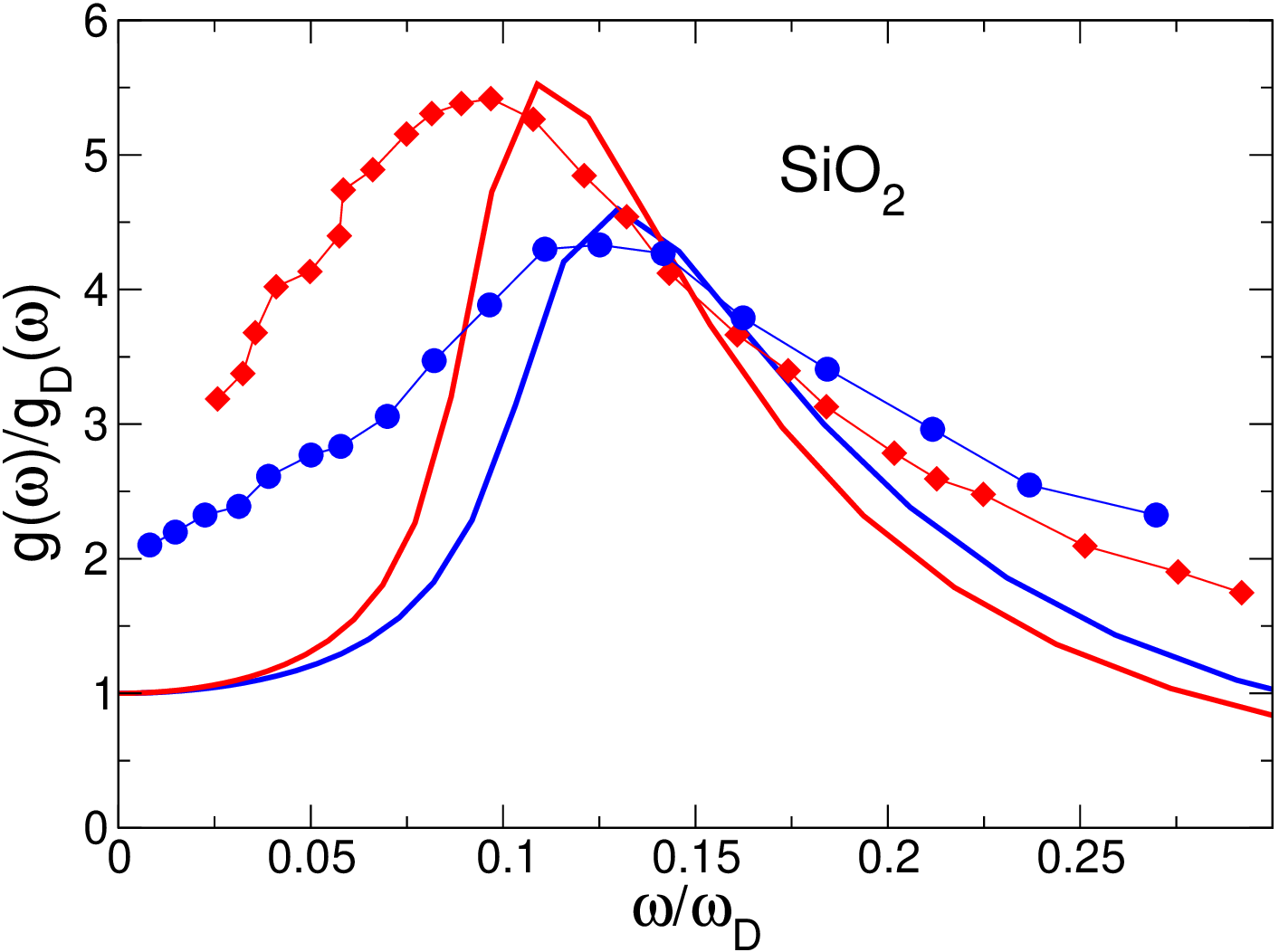}
\caption{\label{cpavekt} Boson peak data for SiO$_2$ \cite{baldi10}
compared with the reduced DOS for the inverse-power model
with two different lower cutoffs $\mu$.}
\end{center}
\end{figure}

In particular, for non-Gaussian distributions of the shear modulus
the quantity $g(\omega)/g_D(\omega)$ can have boson peaks with
arbitrary heights, and a scaling as depicted in Fig. \ref{scalingplot}
holds \cite{koehler11}. As an example we show 
in Fig. \ref{cpavekt} the reduced density
of states, extracted from inelastic X-ray measurements by Baldi
{\it et al.} \cite{baldi10} together with the corresponding
quantity calculated for the inverse-power distribution $P(G)$
given by Eq. (\ref{inverse})

The important difference between the scalar model and the
vector theory (heterogeneous elasticity theory) is
that it describes the physically relevant vector displacements, in
which dilatational and shear degrees of freedom can be distinguished.
Model II. represents a theory, in which the disorder-induced anomalous
frequency dependence is dominated by the dilatation-free shear
degrees of freedom in agreement with recent computer simulations
\cite{marruzzo_forthcoming1,marruzzo_forthcoming2,MonacoSim,shintani08}.

\section{Conclusions}

We have derived a version of the coherent-potential approximation
for both diffusional and vibrational motion in a quenched-disordered
environment, which is 
suitable for topologically disordered materials. The
effective medium is not a crystalline lattice but a homogeneous
and isotropic system with frequency-dependent diffusivity
or elastic constants, resp. . The results can be directly applied
to experimentally measured spectra. In the weak-disorder limit
the CPA has been shown to reduce to the self-consistent Born
approximation, which is based on Gaussian disorder. In the strong-disorder
limit, in which the local diffusivities or elastic quantities vary
exponentially, the CPA has been show to correctly desribe the
percolative aspects of such systems. The disorder-induced vibrational
anomalies have been shown to become stronger as the disorder is increased.
In particular the height of the boson peak has been shown to 
increase indefinitely with the disorder.

In contrast to earlier effective medium theories for the diffusion
problem the present theory includes the correct low-frequency
non-analyticity, which leads to a long-time tail of the
velocity autocorrelation function of the diffusion problem
and to Rayleigh scattering in the vibrational problem.
\section*{Acknowledgement}
S. K. and W. S. are grateful to Prof. Friederike Schmid for helpful discussions.
\section*{Appendix:Electrons in a random potential}

The method developed above for the diffusion and scalar phonon problem can, with slight modifications, be applied to the problem of an electron gas in a random potential. This model is governed by the Schr\"{o}dinger equation 
\begin{eqnarray}
&\int d^3x\,\Psi^\dagger(\bm{x})\left(\frac{\hbar^2}{2m}\Delta-V(\bm{x})-\widetilde{E}\right)\Psi(\bm{x})|\varphi>\nonumber={\bf A}|\varphi>=0&\\&\widetilde{E}=\frac{E+i\varepsilon}{N}&
\end{eqnarray}

The corresponding Green's function can then be expressed with a coherent-state path integral over the Grassmann fields $\theta$ and $\bar{\theta}$:
\begin{eqnarray}
G(\bm{x},\bm{x}^\prime,\widetilde{E})&=&\frac{1}{\zeta[0]}\left.\Fderiv{\zeta[J]}{J(\bm{x}^\prime,\bm{x})}\right|_{J=0}\\
\zeta[J]&=&\int\mathcal{D}[\theta,\bar{\theta}]\,e^{<\theta|A+J|\theta>}
\end{eqnarray}

From here on all steps are analogous to the previous section. First the replica trick is performed and a coarse-grained potential
\eq{
V(\bm{x})=\sum_i v_i\chi_i(\bm{x})
}{}
with no correlation between the coarse-graining boxes is introduced. 
$V(x)$ is subsequently replaced in $A$ 
via the Fadeev-Popov precedure by an auxiliary field $Q$ yielding

\eq{
\langle \zeta^n[J]\rangle=\int \mathcal{D}[Q,\Lambda]\,e^{-nS_{\text{eff}[Q,\Lambda,J]}}
}{electron_RepGenFunc}

with the effective action

\eq{
S_{\text{eff}}[Q,\Lambda,J]=-\tr\ln(A+J)-\sum_i \ln\left\langle e^{V_c/V\Lambda_i(v_i-Q_i)}\right\rangle_i
}{electron_Seff}

As described in the body of the paper, the CPA equations determine the saddle point of (\ref{electron_RepGenFunc}). The saddle point equations read:

\begin{subequations}\label{Electron_finalSaddle}
 \begin{eqnarray}
\Lambda^\prime&=&-\frac{V_c}{V \widetilde{\nu}}\sum_{\bm{k}}\frac{1}{\widetilde{E}-\frac{\hbar^2}{2m} k^2-Q} \label{Electron_L}\\
Q&=&\left\langle\frac{v}{1+\widetilde{\nu}(v-Q)\Lambda^\prime}\right\rangle
\end{eqnarray}
\end{subequations}

The $k$-space summation can now be evaluated and the CPA equations solved. From these results the calculation of the density of states can be done.

The effective medium operator that is defined by equation (\ref{Electron_finalSaddle}) is
\eq{
A_{\text{eff}}(\bm{k},\bm{k}^\prime,z)=\left(z-\frac{\hbar^2k^2}{2m}-Q\right)\delta_{\bm{k} \,\bm{k}^\prime}\, .
}{}

This result has already been obtained in an independent discussion of the CPA for electrons in a random potential \cite{ElectronContCPA} and can
be interpreted as the continuum version of the classical
lattice theories, e.g.  \cite{Yonezawa73}.


\begin{thebibliography}{66}
\expandafter\ifx\csname natexlab\endcsname\relax\def\natexlab#1{#1}\fi
\expandafter\ifx\csname bibnamefont\endcsname\relax
  \def\bibnamefont#1{#1}\fi
\expandafter\ifx\csname bibfnamefont\endcsname\relax
  \def\bibfnamefont#1{#1}\fi
\expandafter\ifx\csname citenamefont\endcsname\relax
  \def\citenamefont#1{#1}\fi
\expandafter\ifx\csname url\endcsname\relax
  \def\url#1{\texttt{#1}}\fi
\expandafter\ifx\csname urlprefix\endcsname\relax\def\urlprefix{URL }\fi
\providecommand{\bibinfo}[2]{#2}
\providecommand{\eprint}[2][]{\url{#2}}

\bibitem[{\citenamefont{Velick\'y et~al.}(1968)\citenamefont{Velick\'y,
  Kirkpatrick, and Ehrenreich}}]{velicky68}
\bibinfo{author}{\bibfnamefont{B.}~\bibnamefont{Velick\'y}},
  \bibinfo{author}{\bibfnamefont{S.}~\bibnamefont{Kirkpatrick}},
  \bibnamefont{and}
  \bibinfo{author}{\bibfnamefont{H.}~\bibnamefont{Ehrenreich}},
  \bibinfo{journal}{Phys. Rev.} \textbf{\bibinfo{volume}{175}},
  \bibinfo{pages}{745} (\bibinfo{year}{1968}).

\bibitem[{\citenamefont{Yonezawa and Morigaki}(1973)}]{Yonezawa73}
\bibinfo{author}{\bibfnamefont{F.}~\bibnamefont{Yonezawa}} \bibnamefont{and}
  \bibinfo{author}{\bibfnamefont{K.}~\bibnamefont{Morigaki}},
  \bibinfo{journal}{Suppl. Prog. Theor. Phys.} \textbf{\bibinfo{volume}{53}},
  \bibinfo{pages}{1} (\bibinfo{year}{1973}).

\bibitem[{\citenamefont{Elliott et~al.}(1974)\citenamefont{Elliott, Krumhansl,
  and Leath}}]{Eliot74}
\bibinfo{author}{\bibfnamefont{R.~J.} \bibnamefont{Elliott}},
  \bibinfo{author}{\bibfnamefont{J.~A.} \bibnamefont{Krumhansl}},
  \bibnamefont{and} \bibinfo{author}{\bibfnamefont{P.~L.} \bibnamefont{Leath}},
  \bibinfo{journal}{Rev. Mod. Phys.} \textbf{\bibinfo{volume}{46}},
  \bibinfo{pages}{465} (\bibinfo{year}{1974}).

\bibitem[{\citenamefont{Jani{\v{s}}}(1989)}]{Janis89}
\bibinfo{author}{\bibfnamefont{V.}~\bibnamefont{Jani{\v{s}}}},
  \bibinfo{journal}{Phys. Rev. B} \textbf{\bibinfo{volume}{40}},
  \bibinfo{pages}{11331} (\bibinfo{year}{1989}).

\bibitem[{\citenamefont{Vollhardt}(2010)}]{Vollhardt10}
\bibinfo{author}{\bibfnamefont{D.}~\bibnamefont{Vollhardt}},
  \bibinfo{journal}{arxiv: 1004.5069v3}  (\bibinfo{year}{2010}).

\bibitem[{\citenamefont{Metzner and Vollhardt}(1989)}]{dmt1}
\bibinfo{author}{\bibfnamefont{W.}~\bibnamefont{Metzner}} \bibnamefont{and}
  \bibinfo{author}{\bibfnamefont{D.}~\bibnamefont{Vollhardt}},
  \bibinfo{journal}{Phys. Rev. Lett.} \textbf{\bibinfo{volume}{62}},
  \bibinfo{pages}{324} (\bibinfo{year}{1989}).

\bibitem[{\citenamefont{Jani\v{s}}(1991)}]{dmt2}
\bibinfo{author}{\bibfnamefont{V.}~\bibnamefont{Jani\v{s}}},
  \bibinfo{journal}{Z. Phys B} \textbf{\bibinfo{volume}{83}},
  \bibinfo{pages}{227} (\bibinfo{year}{1991}).

\bibitem[{\citenamefont{Georges and Kotliar}(1992)}]{dmt3}
\bibinfo{author}{\bibfnamefont{A.}~\bibnamefont{Georges}} \bibnamefont{and}
  \bibinfo{author}{\bibfnamefont{G.}~\bibnamefont{Kotliar}},
  \bibinfo{journal}{Phys. Rev. B} \textbf{\bibinfo{volume}{45}},
  \bibinfo{pages}{6479} (\bibinfo{year}{1992}).

\bibitem[{\citenamefont{Kakehashi}(2002)}]{kakehashi02}
\bibinfo{author}{\bibfnamefont{Y.}~\bibnamefont{Kakehashi}},
  \bibinfo{journal}{Phys. Rev. B} \textbf{\bibinfo{volume}{66}},
  \bibinfo{pages}{104428} (\bibinfo{year}{2002}).

\bibitem[{\citenamefont{Summerfield}(1981)}]{summerfield81}
\bibinfo{author}{\bibfnamefont{S.}~\bibnamefont{Summerfield}},
  \bibinfo{journal}{Sol. State Commun.} \textbf{\bibinfo{volume}{39}},
  \bibinfo{pages}{401} (\bibinfo{year}{1981}).

\bibitem[{\citenamefont{Webman}(1981)}]{webman81}
\bibinfo{author}{\bibfnamefont{I.}~\bibnamefont{Webman}},
  \bibinfo{journal}{Phys. Rev. Lett.} \textbf{\bibinfo{volume}{47}},
  \bibinfo{pages}{1496} (\bibinfo{year}{1981}).

\bibitem[{\citenamefont{Odagaki and Lax}(1981)}]{odagakilax81}
\bibinfo{author}{\bibfnamefont{T.}~\bibnamefont{Odagaki}} \bibnamefont{and}
  \bibinfo{author}{\bibfnamefont{M.}~\bibnamefont{Lax}},
  \bibinfo{journal}{Phys. Rev. B} \textbf{\bibinfo{volume}{24}},
  \bibinfo{pages}{5284} (\bibinfo{year}{1981}).

\bibitem[{\citenamefont{Schirmacher et~al.}(1998)\citenamefont{Schirmacher,
  Diezemann, and Ganter}}]{schirm98}
\bibinfo{author}{\bibfnamefont{W.}~\bibnamefont{Schirmacher}},
  \bibinfo{author}{\bibfnamefont{G.}~\bibnamefont{Diezemann}},
  \bibnamefont{and} \bibinfo{author}{\bibfnamefont{C.}~\bibnamefont{Ganter}},
  \bibinfo{journal}{Phys. Rev. Lett.} \textbf{\bibinfo{volume}{81}},
  \bibinfo{pages}{136} (\bibinfo{year}{1998}).

\bibitem[{\citenamefont{Taraskin et~al.}(2001)\citenamefont{Taraskin, Loh,
  Natarajan, and Elliott}}]{taraskin01}
\bibinfo{author}{\bibfnamefont{S.~N.} \bibnamefont{Taraskin}},
  \bibinfo{author}{\bibfnamefont{Y.~L.} \bibnamefont{Loh}},
  \bibinfo{author}{\bibfnamefont{G.}~\bibnamefont{Natarajan}},
  \bibnamefont{and} \bibinfo{author}{\bibfnamefont{S.~R.}
  \bibnamefont{Elliott}}, \bibinfo{journal}{Phys. Rev. Lett.}
  \textbf{\bibinfo{volume}{86}}, \bibinfo{pages}{1255} (\bibinfo{year}{2001}).

\bibitem[{\citenamefont{Schirmacher}(2006)}]{schirm06}
\bibinfo{author}{\bibfnamefont{W.}~\bibnamefont{Schirmacher}},
  \bibinfo{journal}{Europhys. Lett.} \textbf{\bibinfo{volume}{73}},
  \bibinfo{pages}{892} (\bibinfo{year}{2006}).

\bibitem[{\citenamefont{Schirmacher et~al.}(2007)\citenamefont{Schirmacher,
  Ruocco, and Scopigno}}]{schirm07}
\bibinfo{author}{\bibfnamefont{W.}~\bibnamefont{Schirmacher}},
  \bibinfo{author}{\bibfnamefont{G.}~\bibnamefont{Ruocco}}, \bibnamefont{and}
  \bibinfo{author}{\bibfnamefont{T.}~\bibnamefont{Scopigno}},
  \bibinfo{journal}{Phys. Rev. Lett.} \textbf{\bibinfo{volume}{98}}
  (\bibinfo{year}{2007}).

\bibitem[{\citenamefont{Marruzzo
  et~al.}(2013{\natexlab{a}})\citenamefont{Marruzzo, K\"ohler, Fratalocchi,
  Ruocco, and Schirmacher}}]{marruzzo_forthcoming1}
\bibinfo{author}{\bibfnamefont{A.}~\bibnamefont{Marruzzo}},
  \bibinfo{author}{\bibfnamefont{S.}~\bibnamefont{K\"ohler}},
  \bibinfo{author}{\bibfnamefont{A.}~\bibnamefont{Fratalocchi}},
  \bibinfo{author}{\bibfnamefont{G.}~\bibnamefont{Ruocco}}, \bibnamefont{and}
  \bibinfo{author}{\bibfnamefont{W.}~\bibnamefont{Schirmacher}},
  \bibinfo{journal}{Eur. Phys. J. Special Topics}
  \textbf{\bibinfo{volume}{216}}, \bibinfo{pages}{83}
  (\bibinfo{year}{2013}{\natexlab{a}}).

\bibitem[{\citenamefont{Marruzzo
  et~al.}(2013{\natexlab{b}})\citenamefont{Marruzzo, Schirmacher, Fratalocchi,
  and Ruocco}}]{marruzzo_forthcoming2}
\bibinfo{author}{\bibfnamefont{A.}~\bibnamefont{Marruzzo}},
  \bibinfo{author}{\bibfnamefont{W.}~\bibnamefont{Schirmacher}},
  \bibinfo{author}{\bibfnamefont{A.}~\bibnamefont{Fratalocchi}},
  \bibnamefont{and} \bibinfo{author}{\bibfnamefont{G.}~\bibnamefont{Ruocco}},
  \bibinfo{journal}{Nature Scientific Reports} \textbf{\bibinfo{volume}{3}},
  \bibinfo{pages}{1407} (\bibinfo{year}{2013}{\natexlab{b}}).

\bibitem[{\citenamefont{Schirmacher}(2011)}]{schirm11}
\bibinfo{author}{\bibfnamefont{W.}~\bibnamefont{Schirmacher}},
  \bibinfo{journal}{J. Noncryst. Sol.} \textbf{\bibinfo{volume}{357}},
  \bibinfo{pages}{542} (\bibinfo{year}{2011}).

\bibitem[{\citenamefont{Schirmacher}(2013)}]{schirm13}
\bibinfo{author}{\bibfnamefont{W.}~\bibnamefont{Schirmacher}},
  \bibinfo{journal}{phys. stat. sol. (b)} \textbf{\bibinfo{volume}{250}},
  \bibinfo{pages}{937} (\bibinfo{year}{2013}).

\bibitem[{\citenamefont{B\"ottger and Bryksin}(1985)}]{bottger81}
\bibinfo{author}{\bibfnamefont{H.}~\bibnamefont{B\"ottger}} \bibnamefont{and}
  \bibinfo{author}{\bibfnamefont{V.~V.} \bibnamefont{Bryksin}},
  \emph{\bibinfo{title}{Hopping Conduction in Solids}}
  (\bibinfo{publisher}{Akademie-Verlag}, \bibinfo{address}{Berlin},
  \bibinfo{year}{1985}).

\bibitem[{\citenamefont{Efros and Shklovski\u{i}}(1984)}]{efros84}
\bibinfo{author}{\bibfnamefont{A.~I.} \bibnamefont{Efros}} \bibnamefont{and}
  \bibinfo{author}{\bibfnamefont{B.~I.} \bibnamefont{Shklovski\u{i}}},
  \emph{\bibinfo{title}{Electronic properties of doped semiconductors}}
  (\bibinfo{publisher}{Springer-Verlag}, \bibinfo{address}{Heidelberg},
  \bibinfo{year}{1984}).

\bibitem[{\citenamefont{Ganter and Schirmacher}(2010)}]{ganter10}
\bibinfo{author}{\bibfnamefont{C.}~\bibnamefont{Ganter}} \bibnamefont{and}
  \bibinfo{author}{\bibfnamefont{W.}~\bibnamefont{Schirmacher}},
  \bibinfo{journal}{Phys. Rev. B} \textbf{\bibinfo{volume}{82}},
  \bibinfo{pages}{094205} (\bibinfo{year}{2010}).

\bibitem[{\citenamefont{{Strutt, Third Baron
  Rayleigh}}(1899)}]{RayleighBlueSky}
\bibinfo{author}{\bibfnamefont{J.}~\bibnamefont{{Strutt, Third Baron
  Rayleigh}}}, \bibinfo{journal}{Philos. Mag.} \textbf{\bibinfo{volume}{47}},
  \bibinfo{pages}{375} (\bibinfo{year}{1899}).

\bibitem[{\citenamefont{Ernst and Weyland}(1971)}]{ernst71}
\bibinfo{author}{\bibfnamefont{M.~H.} \bibnamefont{Ernst}} \bibnamefont{and}
  \bibinfo{author}{\bibfnamefont{A.}~\bibnamefont{Weyland}},
  \bibinfo{journal}{Phys. Lett. A} \textbf{\bibinfo{volume}{34}},
  \bibinfo{pages}{39} (\bibinfo{year}{1971}).

\bibitem[{\citenamefont{Machta et~al.}(1984)\citenamefont{Machta, Ernst, and
  van Beijeren~and}}]{machta84}
\bibinfo{author}{\bibfnamefont{J.}~\bibnamefont{Machta}},
  \bibinfo{author}{\bibfnamefont{M.~H.} \bibnamefont{Ernst}}, \bibnamefont{and}
  \bibinfo{author}{\bibfnamefont{H.}~\bibnamefont{van Beijeren~and}},
  \bibinfo{journal}{J. Statist. Phys.} \textbf{\bibinfo{volume}{34}},
  \bibinfo{pages}{413} (\bibinfo{year}{1984}).

\bibitem[{\citenamefont{Ernst et~al.}(1984)\citenamefont{Ernst, Machta,
  Dorfman, and H.}}]{ernst84}
\bibinfo{author}{\bibfnamefont{M.~H.} \bibnamefont{Ernst}},
  \bibinfo{author}{\bibfnamefont{J.}~\bibnamefont{Machta}},
  \bibinfo{author}{\bibfnamefont{J.~R.} \bibnamefont{Dorfman}},
  \bibnamefont{and} \bibinfo{author}{\bibnamefont{H.}}, \bibinfo{journal}{J.
  Statist. Phys.} \textbf{\bibinfo{volume}{34}}, \bibinfo{pages}{477}
  (\bibinfo{year}{1984}).

\bibitem[{\citenamefont{Ganter and Schirmacher}(2011)}]{ganter11}
\bibinfo{author}{\bibfnamefont{C.}~\bibnamefont{Ganter}} \bibnamefont{and}
  \bibinfo{author}{\bibfnamefont{W.}~\bibnamefont{Schirmacher}},
  \bibinfo{journal}{Philos. Magazine} \textbf{\bibinfo{volume}{91}},
  \bibinfo{pages}{1894} (\bibinfo{year}{2011}).

\bibitem[{\citenamefont{Long}(1982)}]{long82}
\bibinfo{author}{\bibfnamefont{A.~R.} \bibnamefont{Long}},
  \bibinfo{journal}{Adv. Phys.} \textbf{\bibinfo{volume}{31}},
  \bibinfo{pages}{553} (\bibinfo{year}{1982}).

\bibitem[{\citenamefont{Jonscher}(1997)}]{jonscher97}
\bibinfo{author}{\bibfnamefont{A.~K.} \bibnamefont{Jonscher}},
  \bibinfo{journal}{Nature} \textbf{\bibinfo{volume}{267}},
  \bibinfo{pages}{673} (\bibinfo{year}{1997}).

\bibitem[{\citenamefont{Dyre and Schr\o{}der}(2000)}]{dyre00}
\bibinfo{author}{\bibfnamefont{J.~C.} \bibnamefont{Dyre}} \bibnamefont{and}
  \bibinfo{author}{\bibfnamefont{T.~B.} \bibnamefont{Schr\o{}der}},
  \bibinfo{journal}{Rev. Mod. Phys.} \textbf{\bibinfo{volume}{72}},
  \bibinfo{pages}{873} (\bibinfo{year}{2000}).

\bibitem[{\citenamefont{Schirmacher and Wagener}(1992)}]{schiwagener92}
\bibinfo{author}{\bibfnamefont{W.}~\bibnamefont{Schirmacher}} \bibnamefont{and}
  \bibinfo{author}{\bibfnamefont{M.}~\bibnamefont{Wagener}},
  \bibinfo{journal}{Philos. Magazine B} \textbf{\bibinfo{volume}{65}},
  \bibinfo{pages}{607} (\bibinfo{year}{1992}).

\bibitem[{\citenamefont{Schirmacher and Wagener}(1993)}]{schiwagener93}
\bibinfo{author}{\bibfnamefont{W.}~\bibnamefont{Schirmacher}} \bibnamefont{and}
  \bibinfo{author}{\bibfnamefont{M.}~\bibnamefont{Wagener}},
  \bibinfo{journal}{Sol. State Comm.} \textbf{\bibinfo{volume}{86}},
  \bibinfo{pages}{597} (\bibinfo{year}{1993}).

\bibitem[{Note1()}]{Note1}
Note1, \bibinfo{note}{we use the conventional bra-ket formalism of quantum
  mechanics, i.e. $<{\protect \bf r}|u^\alpha >=u^\alpha ({\protect \bf r})$,
  etc. .}

\bibitem[{\citenamefont{John et~al.}(1983)\citenamefont{John, Sompolinky, and
  Stephen}}]{john83}
\bibinfo{author}{\bibfnamefont{S.}~\bibnamefont{John}},
  \bibinfo{author}{\bibfnamefont{H.}~\bibnamefont{Sompolinky}},
  \bibnamefont{and} \bibinfo{author}{\bibfnamefont{M.~J.}
  \bibnamefont{Stephen}}, \bibinfo{journal}{Phys. Rev. B}
  \textbf{\bibinfo{volume}{28}}, \bibinfo{pages}{5592} (\bibinfo{year}{1983}).

\bibitem[{\citenamefont{Belitz and Kirkpatrick}(1997)}]{belitz97}
\bibinfo{author}{\bibfnamefont{D.}~\bibnamefont{Belitz}} \bibnamefont{and}
  \bibinfo{author}{\bibfnamefont{T.~R.} \bibnamefont{Kirkpatrick}},
  \bibinfo{journal}{Phys. Rev. B} \textbf{\bibinfo{volume}{56}},
  \bibinfo{pages}{6513} (\bibinfo{year}{1997}).

\bibitem[{Note2()}]{Note2}
Note2, \bibinfo{note}{we work in $d=3$ dimensions throughout this paper,
  although the analysis is not limited to this dimension.}

\bibitem[{Note3()}]{Note3}
Note3, \bibinfo{note}{if the exponential in the denominator of Eq. (\ref
  {saddleb}) would remain in the numerator and then expanded, we would obtain
  the self-consistent Born approximation (see below)}.

\bibitem[{\citenamefont{Butcher and Summerfield}(1981)}]{butcher}
\bibinfo{author}{\bibfnamefont{P.~N.} \bibnamefont{Butcher}} \bibnamefont{and}
  \bibinfo{author}{\bibfnamefont{S.}~\bibnamefont{Summerfield}},
  \bibinfo{journal}{J. Phys. C: Solid State Phys.}
  \textbf{\bibinfo{volume}{14}}, \bibinfo{pages}{L1099} (\bibinfo{year}{1981}).

\bibitem[{\citenamefont{Movaghar and Schirmacher}(1981)}]{Mov2}
\bibinfo{author}{\bibfnamefont{B.}~\bibnamefont{Movaghar}} \bibnamefont{and}
  \bibinfo{author}{\bibfnamefont{W.}~\bibnamefont{Schirmacher}},
  \bibinfo{journal}{J. Phys. C} \textbf{\bibinfo{volume}{14}},
  \bibinfo{pages}{859} (\bibinfo{year}{1981}).

\bibitem[{\citenamefont{Schirmacher et~al.}(2002)\citenamefont{Schirmacher,
  P\"ohlmann, and Maurer}}]{maurer02}
\bibinfo{author}{\bibfnamefont{W.}~\bibnamefont{Schirmacher}},
  \bibinfo{author}{\bibfnamefont{M.}~\bibnamefont{P\"ohlmann}},
  \bibnamefont{and} \bibinfo{author}{\bibfnamefont{E.}~\bibnamefont{Maurer}},
  \bibinfo{journal}{phys. stat. sol. (b)} \textbf{\bibinfo{volume}{230}},
  \bibinfo{pages}{31} (\bibinfo{year}{2002}).

\bibitem[{\citenamefont{Schirmacher et~al.}(2004)\citenamefont{Schirmacher,
  Maurer, and P\"ohlmann}}]{maurer04}
\bibinfo{author}{\bibfnamefont{W.}~\bibnamefont{Schirmacher}},
  \bibinfo{author}{\bibfnamefont{E.}~\bibnamefont{Maurer}}, \bibnamefont{and}
  \bibinfo{author}{\bibfnamefont{M.}~\bibnamefont{P\"ohlmann}},
  \bibinfo{journal}{phys. stat. sol. (c)} \textbf{\bibinfo{volume}{1}},
  \bibinfo{pages}{17} (\bibinfo{year}{2004}).

\bibitem[{\citenamefont{Maurer and Schirmacher}(2004)}]{MaurerFELO}
\bibinfo{author}{\bibfnamefont{E.}~\bibnamefont{Maurer}} \bibnamefont{and}
  \bibinfo{author}{\bibfnamefont{W.}~\bibnamefont{Schirmacher}},
  \bibinfo{journal}{J. Low Temp. Phys.} \textbf{\bibinfo{volume}{137}},
  \bibinfo{pages}{453} (\bibinfo{year}{2004}).

\bibitem[{\citenamefont{Bruggeman}(1935)}]{bruggemann35}
\bibinfo{author}{\bibfnamefont{D.}~\bibnamefont{Bruggeman}},
  \bibinfo{journal}{Ann. Phys} \textbf{\bibinfo{volume}{416}},
  \bibinfo{pages}{636} (\bibinfo{year}{1935}).

\bibitem[{\citenamefont{Landauer}(1952)}]{landauer52}
\bibinfo{author}{\bibfnamefont{R.}~\bibnamefont{Landauer}},
  \bibinfo{journal}{J. Appl. Phys.} \textbf{\bibinfo{volume}{23}},
  \bibinfo{pages}{779} (\bibinfo{year}{1952}).

\bibitem[{\citenamefont{Kirkpatrick}(1973)}]{kirkpatrick73}
\bibinfo{author}{\bibfnamefont{S.}~\bibnamefont{Kirkpatrick}},
  \bibinfo{journal}{Rev. Mod. Phys.} \textbf{\bibinfo{volume}{45}},
  \bibinfo{pages}{574} (\bibinfo{year}{1973}).

\bibitem[{\citenamefont{Dyre}(1994)}]{dyre94}
\bibinfo{author}{\bibfnamefont{J.~C.} \bibnamefont{Dyre}},
  \bibinfo{journal}{Phys. Rev. B} \textbf{\bibinfo{volume}{49}},
  \bibinfo{pages}{11709} (\bibinfo{year}{1994}).

\bibitem[{\citenamefont{Gochanour et~al.}(1979)\citenamefont{Gochanour,
  Andersen, and Fayer}}]{Andersen}
\bibinfo{author}{\bibfnamefont{C.}~\bibnamefont{Gochanour}},
  \bibinfo{author}{\bibfnamefont{H.~C.} \bibnamefont{Andersen}},
  \bibnamefont{and} \bibinfo{author}{\bibfnamefont{M.}~\bibnamefont{Fayer}},
  \bibinfo{journal}{J. Chem. Phys.} \textbf{\bibinfo{volume}{70}},
  \bibinfo{pages}{4254} (\bibinfo{year}{1979}).

\bibitem[{\citenamefont{H\"ofling and Franosch}(2007)}]{hofling07}
\bibinfo{author}{\bibfnamefont{F.}~\bibnamefont{H\"ofling}} \bibnamefont{and}
  \bibinfo{author}{\bibfnamefont{T.}~\bibnamefont{Franosch}},
  \bibinfo{journal}{Phys. Rev. Lett.} \textbf{\bibinfo{volume}{98}},
  \bibinfo{pages}{140601} (\bibinfo{year}{2007}).

\bibitem[{\citenamefont{Spanner et~al.}(2011)\citenamefont{Spanner, H\"ofling,
  Schr\"oder-Turk, Mecke, and Franosch}}]{spanner11}
\bibinfo{author}{\bibfnamefont{M.}~\bibnamefont{Spanner}},
  \bibinfo{author}{\bibfnamefont{F.}~\bibnamefont{H\"ofling}},
  \bibinfo{author}{\bibfnamefont{G.~E.} \bibnamefont{Schr\"oder-Turk}},
  \bibinfo{author}{\bibfnamefont{K.}~\bibnamefont{Mecke}}, \bibnamefont{and}
  \bibinfo{author}{\bibfnamefont{T.}~\bibnamefont{Franosch}},
  \bibinfo{journal}{J. Phys.: Condens. Matter} \textbf{\bibinfo{volume}{23}},
  \bibinfo{pages}{234120} (\bibinfo{year}{2011}).

\bibitem[{\citenamefont{Schirmacher}(1988)}]{SolidStateIonics}
\bibinfo{author}{\bibfnamefont{W.}~\bibnamefont{Schirmacher}},
  \bibinfo{journal}{Sol. State ionics} \textbf{\bibinfo{volume}{28-30}},
  \bibinfo{pages}{129} (\bibinfo{year}{1988}).

\bibitem[{\citenamefont{Schr\o{}der and Dyre}(2000{\natexlab{a}})}]{dyre08}
\bibinfo{author}{\bibfnamefont{T.~B.} \bibnamefont{Schr\o{}der}}
  \bibnamefont{and} \bibinfo{author}{\bibfnamefont{J.~C.} \bibnamefont{Dyre}},
  \bibinfo{journal}{Phys. Rev. Lett.} \textbf{\bibinfo{volume}{101}},
  \bibinfo{pages}{025901} (\bibinfo{year}{2000}{\natexlab{a}}).

\bibitem[{\citenamefont{Cottaar et~al.}(2011)\citenamefont{Cottaar, Koster,
  Coehoorn, and Bobbert}}]{bobbert}
\bibinfo{author}{\bibfnamefont{J.}~\bibnamefont{Cottaar}},
  \bibinfo{author}{\bibfnamefont{L.~J.~A.} \bibnamefont{Koster}},
  \bibinfo{author}{\bibfnamefont{R.}~\bibnamefont{Coehoorn}}, \bibnamefont{and}
  \bibinfo{author}{\bibfnamefont{P.~A.} \bibnamefont{Bobbert}},
  \bibinfo{journal}{Phys. Rev. Lett.} \textbf{\bibinfo{volume}{107}},
  \bibinfo{pages}{136601} (\bibinfo{year}{2011}).

\bibitem[{\citenamefont{Mott and Davis}(1971)}]{mott}
\bibinfo{author}{\bibfnamefont{N.~F.} \bibnamefont{Mott}} \bibnamefont{and}
  \bibinfo{author}{\bibfnamefont{E.~A.} \bibnamefont{Davis}},
  \emph{\bibinfo{title}{Electronic Processes in Non-Crystalline Materials}}
  (\bibinfo{publisher}{Clarendon}, \bibinfo{address}{Oxford},
  \bibinfo{year}{1971}).

\bibitem[{\citenamefont{Wong and Angell}(1976)}]{WongAngel}
\bibinfo{author}{\bibfnamefont{J.}~\bibnamefont{Wong}} \bibnamefont{and}
  \bibinfo{author}{\bibfnamefont{C.~A.} \bibnamefont{Angell}},
  \emph{\bibinfo{title}{Glass : structure by spectroscopy}}
  (\bibinfo{publisher}{M. Dekker}, \bibinfo{year}{1976}), ISBN
  \bibinfo{isbn}{0824764684}.

\bibitem[{Note4()}]{Note4}
Note4, \bibinfo{note}{note that $E_a$ is {\protect \it not} equal to the
  maximum barrier height $E^*$ but much smaller and is determined by the self
  consistent Eq. (\ref {barr}).}

\bibitem[{\citenamefont{Dyre}(1985)}]{dyre85}
\bibinfo{author}{\bibfnamefont{J.~C.} \bibnamefont{Dyre}},
  \bibinfo{journal}{Phys. Lett.} \textbf{\bibinfo{volume}{108A}},
  \bibinfo{pages}{457} (\bibinfo{year}{1985}).

\bibitem[{\citenamefont{Schr\o{}der and Dyre}(2000{\natexlab{b}})}]{dyre00a}
\bibinfo{author}{\bibfnamefont{T.~B.} \bibnamefont{Schr\o{}der}}
  \bibnamefont{and} \bibinfo{author}{\bibfnamefont{J.~C.} \bibnamefont{Dyre}},
  \bibinfo{journal}{Phys. Rev. Lett.} \textbf{\bibinfo{volume}{84}},
  \bibinfo{pages}{310} (\bibinfo{year}{2000}{\natexlab{b}}).

\bibitem[{\citenamefont{Long et~al.}(1988)\citenamefont{Long, Mcmillan, Balkan,
  and Summerfield}}]{Lossfunction}
\bibinfo{author}{\bibfnamefont{A.}~\bibnamefont{Long}},
  \bibinfo{author}{\bibfnamefont{J.}~\bibnamefont{Mcmillan}},
  \bibinfo{author}{\bibfnamefont{N.}~\bibnamefont{Balkan}}, \bibnamefont{and}
  \bibinfo{author}{\bibfnamefont{S.}~\bibnamefont{Summerfield}},
  \bibinfo{journal}{Phil. Mag. B} \textbf{\bibinfo{volume}{58}},
  \bibinfo{pages}{153} (\bibinfo{year}{1988}).

\bibitem[{\citenamefont{Pasveer et~al.}(2006)\citenamefont{Pasveer, Bobbert,
  and Michels}}]{pasveer06}
\bibinfo{author}{\bibfnamefont{W.~F.} \bibnamefont{Pasveer}},
  \bibinfo{author}{\bibfnamefont{P.~A.} \bibnamefont{Bobbert}},
  \bibnamefont{and} \bibinfo{author}{\bibfnamefont{M.~A.~J.}
  \bibnamefont{Michels}}, \bibinfo{journal}{Phys. Rev. B}
  \textbf{\bibinfo{volume}{74}}, \bibinfo{pages}{165209}
  (\bibinfo{year}{2006}).

\bibitem[{\citenamefont{Landau and Lifshitz}(1959)}]{LandauLifshitz}
\bibinfo{author}{\bibfnamefont{L.}~\bibnamefont{Landau}} \bibnamefont{and}
  \bibinfo{author}{\bibfnamefont{E.}~\bibnamefont{Lifshitz}},
  \emph{\bibinfo{title}{Theory of Elasticity}} (\bibinfo{publisher}{Pergamon
  Press}, \bibinfo{year}{1959}).

\bibitem[{\citenamefont{K\"{o}hler}(2011)}]{koehler11}
\bibinfo{author}{\bibfnamefont{S.}~\bibnamefont{K\"{o}hler}},
  \bibinfo{type}{Diploma thesis}, \bibinfo{school}{Universit\"{a}t Mainz}
  (\bibinfo{year}{2011}).

\bibitem[{\citenamefont{Baldi et~al.}(2010)\citenamefont{Baldi, Giordano,
  Monaco, and Ruta}}]{baldi10}
\bibinfo{author}{\bibfnamefont{G.}~\bibnamefont{Baldi}},
  \bibinfo{author}{\bibfnamefont{V.~M.} \bibnamefont{Giordano}},
  \bibinfo{author}{\bibfnamefont{G.}~\bibnamefont{Monaco}}, \bibnamefont{and}
  \bibinfo{author}{\bibfnamefont{B.}~\bibnamefont{Ruta}},
  \bibinfo{journal}{Phys. Rev. Lett.} \textbf{\bibinfo{volume}{104}},
  \bibinfo{pages}{195501} (\bibinfo{year}{2010}).

\bibitem[{\citenamefont{Monaco and Mossa}(2009)}]{MonacoSim}
\bibinfo{author}{\bibfnamefont{G.}~\bibnamefont{Monaco}} \bibnamefont{and}
  \bibinfo{author}{\bibfnamefont{S.}~\bibnamefont{Mossa}},
  \bibinfo{journal}{PNAS} \textbf{\bibinfo{volume}{106}},
  \bibinfo{pages}{16907} (\bibinfo{year}{2009}).

\bibitem[{\citenamefont{Shintani and Tanaka}(2008)}]{shintani08}
\bibinfo{author}{\bibfnamefont{H.}~\bibnamefont{Shintani}} \bibnamefont{and}
  \bibinfo{author}{\bibfnamefont{H.}~\bibnamefont{Tanaka}},
  \bibinfo{journal}{Nature Materials} \textbf{\bibinfo{volume}{7}},
  \bibinfo{pages}{870} (\bibinfo{year}{2008}).

\bibitem[{\citenamefont{Zimmermann and Schindler}(2009)}]{ElectronContCPA}
\bibinfo{author}{\bibfnamefont{R.}~\bibnamefont{Zimmermann}} \bibnamefont{and}
  \bibinfo{author}{\bibfnamefont{C.}~\bibnamefont{Schindler}},
  \bibinfo{journal}{Phys. Rev. B} \textbf{\bibinfo{volume}{80}},
  \bibinfo{pages}{144202} (\bibinfo{year}{2009}).

\end{thebibliography}
\end{document}